\documentclass[aps,prb,twocolumn,floats,epsfig,pdflatex]{revtex4}
\UseRawInputEncoding
\usepackage{amssymb,amsbsy,amsmath,mathrsfs}
\usepackage{epsfig,psfrag,subfigure}
\usepackage{graphicx,float}
\usepackage{times}
\usepackage{color}
\usepackage{subfigure}
\usepackage{setspace}
\usepackage{bm} 
\usepackage{caption}

\newcommand\bea{\begin{eqnarray}}
\newcommand\eea{\end{eqnarray}}
\newcommand\beq{\begin{equation}}
\newcommand\eeq{\end{equation}}
\newcommand\bib{\bibitem}

\newcommand{\noi}{\noindent}
\newcommand{\non}{\nonumber}
\newcommand{\al}{\alpha}
\newcommand{\de}{\delta}

\newcommand{\lm}{\lambda}

\newcommand{\da}{\dagger}
\newcommand{\pa}{\partial}

\newcommand{\bra}[1]{\langle #1|}
\newcommand{\ket}[1]{|#1\rangle}

\begin{document}

\title{Bosonization study of a generalized statistics model with four Fermi
points}
\author{Sreemayee Aditya$^1$ and Diptiman Sen$^{1,2}$}
\affiliation{$^1$Center for High Energy Physics, Indian Institute of Science,
Bengaluru 560012, India \\
$^2$Department of Physics, Indian Institute of Science, Bengaluru 560012,
India}

\begin{abstract}
We study a one-dimensional lattice model of generalized statistics in which 
particles have next-nearest-neighbor hopping between sites which depends 
on the occupation number
at the intermediate site and a statistical parameter $\phi$. The model breaks 
parity and time-reversal symmetries and has four-fermion interactions if 
$\phi \ne 0$. We first analyze the model using mean field theory and find that 
there are four Fermi points whose locations depend on $\phi$ and the filling 
$\eta$. We then study the modes near the Fermi points using the technique of 
bosonization. Based on the quadratic terms in the bosonized Hamiltonian, we 
find that the low-energy modes form two decoupled Tomonaga-Luttinger liquids 
with different values of the Luttinger parameters which depend on $\phi$ and 
$\eta$; further, the right and left moving modes of each system have different 
velocities. A study of the scaling dimensions of the cosine terms in the 
Hamiltonian indicates that the terms appearing in one of the Tomonaga-Luttinger
liquids will flow under the renormalization group and the system may reach a 
nontrivial fixed point in the long distance limit. We examine the scaling 
dimensions of 
various charge density and superconducting order parameters to find which of 
them is the most relevant for different values of $\phi$ and $\eta$. Finally 
we look at two-particle bound states that appear in this system and discuss 
their possible relevance to the properties of the system in the thermodynamic 
limit. Our work shows that the low-energy properties of this model of 
generalized statistics have a rich structure as a function of $\phi$ and $\eta$.
\end{abstract}

\maketitle

\section{Introduction}
\label{sec1}

The possibility of identical particles having generalized statistics in one 
dimension has been studied extensively over many years. Such generalizations
can be introduced in many different ways, for instance, by modifying the
conditions on the wave function and its derivative at the points when
two of the particles have the same coordinate, modifying the
commutation relations between the creation and annihilation operators
in a second-quantized formalism, or modifying the form of the exclusion 
principle~\cite{lieb,haldane,bal,wu,ha,murthy,rabello,kundu,batch,hao,posske,
poly,agarwala1,bonkhoff,santra}. Several theoretical proposals have been 
made for realizing generalized statistics in one 
dimension~\cite{keil,strater,gres,card,gres2}.

A recent paper has studied a model of pseudofermions on a 
one-dimensional lattice in which the second quantized operators have a
generalized statistics governed by a parameter $\phi$~\cite{agarwala1}. 
The model has both nearest- and next-nearest-neighbor hoppings $t_1$ and 
$t_2$, and the latter is sensitive to $\phi$. At half-filling, it has a 
rich phase diagram as a function of $t_1/t_2$ and $\phi$. The model has 
two Fermi points when $|t_1 / t_2 | > 2$ and four Fermi points when 
$|t _1 / t_2| < 2$, with a Lifshitz transition occurring between the 
two phases at $|t_1 /t_2| = 2$. The phase with two Fermi points has been 
studied in detail using bosonization~\cite{agarwala1}. However, the
phase with four Fermi points is more difficult to study as it requires
the diagonalization of a model with two right-moving and two left-moving 
modes. In this paper, we aim to analyze this phase in detail for arbitrary 
values of the filling.

The plan of this paper is as follows. In Sec.~\ref{sec2}, we introduce
our model of pseudofermions with generalized statistics. In order to focus
on the phase with four Fermi points, we consider only next-nearest-neighbor
hoppings which have a phase which depends on the particle number on the
intermediate site and a parameter $\phi$. In Sec.~\ref{sec3}, we analyze the 
model using mean field theory. This enables us to find the locations of
the four Fermi points as a function of $\phi$ and the filling which is 
governed by a parameter $\eta$. In Sec.~\ref{sec4}, we use the bosonization 
method to study the modes close to the Fermi points. We find that the
bosonized Hamiltonian has terms which are quadratic in the bosonic fields
and terms which involve cosines of those fields. We diagonalize the 
quadratic part of the Hamiltonian, thereby finding that the model consists
of two decoupled Tomonaga-Luttinger liquids with separate Luttinger
parameters $K_1$ and $K_2$ and velocities of right- and left-moving 
modes. We then calculate the scaling dimensions of the cosine terms and 
discuss what these may imply about the long-distance properties of the model.
We also find the scaling dimensions of various charge density and 
superconducting order parameters to determine which of them is likely to
dominate the long-distance properties. In Sec.~\ref{sec5}, we study a
system with only two particles and show that this has both continuum
and bound states. We examine the implication of the bound states for the
properties of the system with a large number of particles. In Sec.~\ref{sec6},
we summarize our results, point out some directions for future studies,
and mention possible realizations of our model. In the Appendices we 
discuss some technical details like a Bogoliubov transformation for 
bosonic fields with unequal right- and left-moving velocities and a 
nonlocal mapping between models with $\phi$ and $\pi + \phi$.

\section{Generalized statistics in one dimension, Hamiltonian and symmetries}
\label{sec2}

In this section, we will study a lattice model for generalized statistics 
which was introduced in Ref.~\onlinecite{agarwala1}.
The generalized algebra of creation and annihilation operators 
of pseudofermions on sites $j$ and $k$ is given by
\begin{eqnarray} a_j a_{k} ~+~ a_k a_j e^{i\phi \text{sgn}(k-j)} &=& 0, \non \\
a_j a_k^\da ~+~ a_k^\da a_j e^{-i\phi \text{sgn}(k-j)} &=& \delta_{jk}, \non \\
{[} N_j, a_k {]} &=& - ~\delta_{jk} ~a_{k}, \non \\
{[} N_j, a_k^\da {]} &=& \delta_{jk} ~a_{k}^\da, \label{stat} \end{eqnarray}
where $N_{j}=a_{j}^{\da}a_{j}$ is the occupation number of pseudofermions 
on site $j$. The definition $\text{sgn} (0)=0$ generates the 
algebra of pseudofermions for $j=k$ which is consistent with the algebra of 
usual fermions. In contrast, the algebra for $j \ne k$ is different 
and it can be tuned from ordinary fermions to hard core bosons by tuning the 
statistical phase from $\phi=0$ to $\phi=\pi$.

It is clear from Eq.~\eqref{stat} that changing $\phi \to \phi + 2 \pi$
makes no difference. Hence it is enough to study values of $\phi$ lying in 
the range $[-\pi,\pi]$. We also see that the system remains unchanged if we 
change $\phi \to - \phi$ and do a parity transformation $j \to - j$ for all 
$j$. We will therefore only consider the range $0 \le \phi \le \pi$ in 
this paper.

\subsection{Hamiltonian}
\label{sec2a}

We consider the following Hamiltonian for a model of pseudofermions, 
\begin{equation}
H=-\sum_{j} ~[(t_{1}a_{j}^{\da}a_{j+1}+t_{2}a_{j}^{\da}a_{j+2}
+\text{H.c.})+\mu a_{j}^{\da}a_{j}], \label{ham1} \end{equation}
where $t_{1}$ and $t_{2}$ are the nearest- and next-nearest-neighbor hopping 
amplitudes respectively, and $\mu$ is the chemical potential. 
(Throughout this paper we will set both $\hbar$ and the lattice spacing $a$ 
to unity). We can map this to a Hamiltonian of ordinary 
(spinless) fermions by the fractional Jordan-Wigner transformation,
\begin{eqnarray}
c_{j} &=& K_{j}a_{j}, ~~~~ c_{j}^{\da}=a_{j}^{\da}K_{j}^\da, \non \\
\text{where} ~~~K_{j} &=& e^{-i\phi\sum_{k<j}n_{k}}, \end{eqnarray}
and $c_{j}$ and $c_{j}^{\da}$ are the creation and annihilation operators 
of fermions with the usual anticommuting algebra. Equation~\eqref{ham1} is 
then mapped into a Hamiltonian of ordinary fermions
\beq H=-\sum_{j} ~[(t_{1}c_j^{\da}c_{j+1}+t_{2}e^{i\phi n_{j+1}} c_j^{\da}
c_{j+2}+\text{H.c.}) +\mu c_j^{\da}c_{j}]. \label{ham2} \eeq
where $n_{j}=c_{j}^{\da}c_{j}$. We note that the number operator $N_j$ for
pseudofermions is mapped to the number operator of the ordinary fermions 
$n_{j}$ by this transformation. The first term, nearest-neighbor hopping
with amplitude $t_1$, remains unaffected by the phase $\phi$. However, 
the next-nearest-neighbor hopping carries the information of the statistical 
phase: the fermions hop with a phase which 0 if the intermediate phase
is empty and is $\pm \phi$ if the intermediate site is filled and the 
hopping is to occurs to the left (right) respectively. This 
dependence of the next-nearest-neighbor hopping on the statistical phase 
$\phi$ makes it evident that a nonzero finite $t_{2}$ is necessary to obtain 
nontrivial phases in this model. This motivates us to explore the limit 
where we have a finite $t_{2}$, but $t_{1}$ is set equal to zero. We will
therefore consider the Hamiltonian
\begin{equation} H=-\sum_{j} ~[ (t_{2}e^{i\phi n_{j+1}}c_{j}^{\da}c_{j+2}+
\text{H.c.}) + \mu c_{j}^{\da}c_{j}]. \label{ham3} \end{equation}
Note that Eq.~\eqref{ham3} remains invariant under $t_2 \to - t_2$ since
we can change the sign of those terms by carrying out the transformation
\beq c_j ~\to~ e^{ij \pi/2} ~c_j ~~~{\rm and}~~~ c_j^\da ~\to~ e^{-ij \pi/2} ~
c_j^\da. \label{transf1} \eeq

\begin{figure}[htb]
\centering
\includegraphics[width=0.45\textwidth]{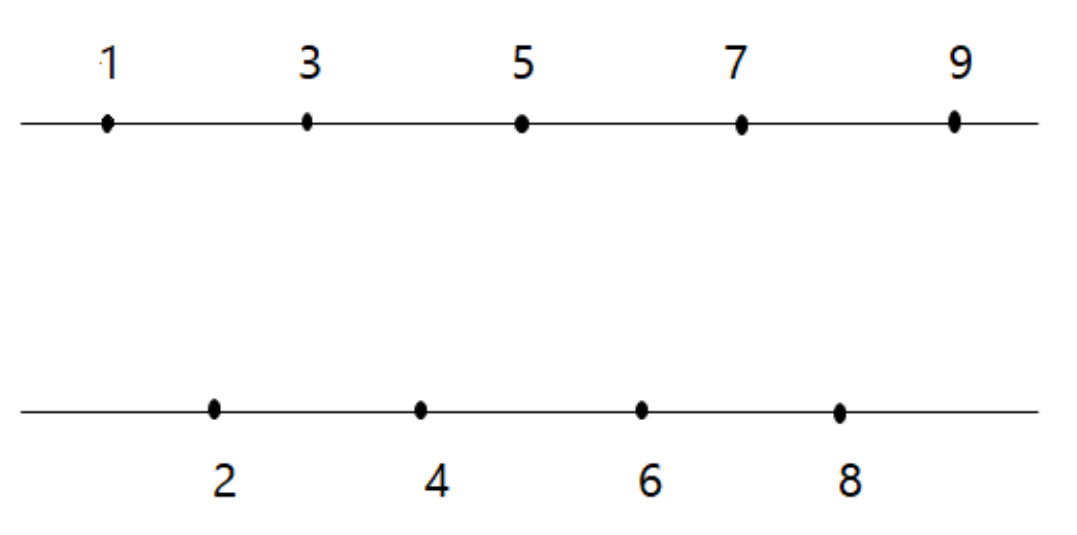}
\caption{Schematic picture of the system showing the two sublattices 
corresponding to odd and even values of the site label.} 
\label{sa1fig01} \end{figure}

A schematic picture of our model is shown in Fig.~\ref{sa1fig01}. Hopping 
only occurs between nearest-neighbor sites on the same sublattice, 
corresponding to either odd or even values of $j$; the hopping amplitude
depends on the occupation number of the intermediate site which belongs to
the other sublattice. We note here that a system with $t_1 = 0$ can be 
physically realized if the sublattices are replaced by the two spin components 
of a spin-1/2 particle; such a system naturally has $t_1 = 0$ if the hopping 
conserves the spin component. A system similar to this has been experimentally 
studied in Ref.~\onlinecite{gorg}.

We will now do the following transformation on the creation and 
annihilation operators,
\begin{equation} c_{j} ~\rightarrow~ c_{j}e^{-ij \phi/4}, ~~~ c_{j}^{\da} ~
\rightarrow~ c_{j}^{\da}e^{ij \phi /4}. \end{equation}
Then Eq.~\eqref{ham3} takes the form 
\begin{equation} H=-\sum_{j} ~[(t_{2}e^{i\phi (n_{j+1}-1/2)}c_{j}^{\da}c_{j+2}+
\text{H.c.})+\mu c_{j}^{\da}c_{j}]. \label{ham4} \end{equation}
This is the Hamiltonian that we will study in the rest of this paper.
The choice of the term $n_{j+1} - 1/2$, rather than just $n_{j+1}$, in 
the phase is motivated by a particle-hole transformation which will be
discussed later. 

Note that the number of particles on each sublattice (either even or 
odd-numbered sites) is a conserved quantity since the hopping only occurs 
within each sublattice separately.

\subsection{Symmetries}
\label{sec2b}

Next, we will the symmetries of our model. First, we examine how the 
Hamiltonian in Eq.~\eqref{ham4} behaves under a particle-hole transformation. 
To discuss that, we add a constant to turn the Hamiltonian into
\begin{eqnarray} H &=&- t_2 \sum_{j} ~[e^{i\phi (n_{j+1}-1/2)}c_{j}^{\da}
c_{j+2}+ \text{H.c.}] \non\\
&&- \mu \sum_{j} ~[c_{j}^{\da}c_{j}-1/2 ]. \label{ham5} \end{eqnarray}
Under a particle-hole transformation, we have 
\begin{eqnarray} c_{j} \rightarrow c_{j}^{\da} ~~~{\rm and}~~~ 
c_{j}^{\da} \rightarrow c_{j}. \end{eqnarray}
As a result of this transformation, we find that $n_j - 1/2 \to - (n_j - 1/2)$.
The Hamiltonian Eq.~\eqref{ham5} then flips sign and we obtain
\begin{eqnarray} H&=& t_2 \sum_{j} ~[e^{i\phi (n_{j+1}-1/2)}c_{j}^{\da}
c_{j+2}+\text{H.c.}] \non \\
&&+ \mu \sum_{j} ~[c_{j}^{\da}c_{j}-1/2 ]. \label{ham6} \end{eqnarray}
We then carry out the transformation given in Eq.~\eqref{transf1} to change
the Hamiltonian in Eq.~\eqref{ham6} to
\begin{eqnarray} H&=&-t_2 \sum_{j} ~[e^{i\phi (n_{j+1}-1/2)}c_{j}^{\da}
c_{j+2}+ \text{H.c.}] \non\\
&&+ \mu \sum_{j} ~[c_{j}^{\da}c_{j}-1/2] \label{ham7}. \end{eqnarray}
Comparing Eqs.~\eqref{ham5} and \eqref{ham7}, we see that the Hamiltonian 
remains invariant under a particle-hole transformation provided that we also 
change $\mu \rightarrow-\mu$. (We note that the Hamiltonian has this 
invariance only if $t_1 = 0$).

We will now discuss parity ($P$) and time-reversal ($T$) transformations. 
Under $P$, the creation and annihilation operators transform as 
\begin{eqnarray} &c_{j}&\:\rightarrow\:c_{-j},\non\\
&c_{j}^{\da}&\:\rightarrow\:c_{-j}^{\da}. \end{eqnarray}
The Hamiltonian in Eq.~\eqref{ham4} then becomes
\begin{eqnarray} H &=&-\sum_{j} ~[t_{2}e^{i\phi (n_{-(j+1)}-1/2)}
c_{-j}^{\da} c_{-(j+2)}+\text{H.c.}) \non \\
&& ~~~~~~~~~~~+\mu c_{-j}^{\da}c_{-j}], \end{eqnarray}
which can be written as
\begin{equation} H = - \sum_{j} ~[ t_2 (e^{i\phi (n_{j+1}-1/2)}c_{j+2}^{\da}
c_{j}+ \text{H.c.}) + \mu c_{j}^{\da}c_{j}]. \label{ham8} \end{equation}
Thus the Hamiltonian in Eq.~\eqref{ham4} is not invariant under 
$P$, unless we also flip $\phi \to - \phi$. 
Similarly, under time-reversal $T$, we complex conjugate the Hamiltonian which 
implies that Eq.~\eqref{ham4} transforms into Eq.~\eqref{ham8}. Hence 
the Hamiltonian is not invariant under $T$. However, the Hamiltonian is 
invariant under $PT$.

\section{Mean field theory}
\label{sec3}

We begin our discussion by considering the special case $\phi=0$ which
describes a system of noninteracting fermions. The energy-momentum 
dispersion is then given by
\begin{equation} E_k ~=~ -2t_{2}\cos(2k) ~-~ \mu, \end{equation}
where $k$ lies in the range $[-\pi,\pi]$. We see that this system has four 
Fermi points if the chemical potential lies in the range $- 2t_2 < \mu < 2t_2$.
For $\mu = 0$, the Fermi points lie at $k = \pm \pi/4$ and $\pm 3 \pi/4$. 

We now discuss a mean field treatment for general $\phi$ to include the 
effect of the statistical interaction. In the rest of this paper we will
set $t_{2}=-1$ for convenience. The Hamiltonian is then
\begin{equation} H=\sum_{j} ~[e^{i\phi (n_{j+1}-1/2)}c_{j}^{\da}c_{j+2}+
\text{H.c.}- \mu c_{j}^{\da}c_{j}]. \label{ham9} \end{equation}
Now, the exponential factor can be written in a more convenient form by 
noting that $n_j$ can only take the values zero or 1. Hence
\begin{eqnarray}
(n_{j}-1/2)^p &=& (1/2)^p, ~~\text{if} ~~p~~ \text{is ~even} \non \\
(n_{j}-1/2)^p &=& (1/2)^{p-1} (n_{j}-1/2) ~~\text{if} ~~p~~ \text{is ~odd}. 
\end{eqnarray}
The phase factor can therefore be written as 
\begin{equation} e^{i\phi (n_{j}-1/2)}=\cos(\phi/2)+2i\sin(\phi/2)(n_{j}-1/2).
\end{equation}
The Hamiltonian in Eq.~\eqref{ham9} can now be written as the sum of a 
noninteracting part $H_0$ (which is quadratic in the fermion operators) and 
an interacting part $H_{int}$ (quartic),
\begin{eqnarray} H &=& H_{0}+H_{int}, \non \\
H_{0} &=& \sum_{j} ~[\cos(\phi/2)(c_{j}^\da c_{j+2}+c_{j+2}^\da c_{j}) \non \\
&& ~~~~~~~-i\sin(\phi/2)(c_{j}^\da c_{j+2}-c_{j+2}^\da c_{j}) -\mu 
c_{j}^\da c_{j}], \non \\
H_{int} &=& 2i\sin(\phi/2)\sum_{j} n_{j+1}(c_{j}^\da c_{j+2}-
c_{j+2}^\da c_{j}). \label{ham10} \end{eqnarray}
Following a mean field treatment, the interacting part becomes
\begin{widetext}
\begin{equation} \sum_{j}n_{j+1}(c_{j}^{\da}c_{j+2}-c_{j+2}^{\da} c_{j})
\rightarrow \sum_j [\langle n_{j+1} \rangle (c_{j}^{\da} c_{j+2}-
c_{j+2}^{\da} c_{j}) + \langle c_{i}^{\da} c_{j+2}-c_{j+2}^{\da}
c_{j} \rangle c_{j}^{\da}c_{j}]. \label{mft} \end{equation}
\end{widetext}
In the first term in Eq.~\eqref{mft}, we replace $\langle n_{j+1} \rangle 
\to 1/2+\eta$, where $\eta$ denotes the deviation from half-filling 
and lies in the range $[-1/2,1/2]$, where $\eta=-1/2$ and $1/2$ correspond 
to a completely empty and completely filled band respectively. The second term 
in Eq.~\eqref{mft} corresponds to a shift in the chemical potential. 
Its effect can be absorbed by introducing a new chemical potential $\mu'$
\begin{equation} 2i\sin(\phi/2)\big<c_{j}^{\da} c_{j+2}-c_{j+2}^{\da}
c_{j}\big>-\mu =-\mu'. \end{equation}
[Note that in the mean field treatment, we are setting expectation values of
nearest-neighbor terms $\langle c_j^\da c_{j\pm 1} \rangle = 0$.
This is because we are analyzing a model with nearest-neighbor hopping
$t_1 = 0$. Hence the number of fermions on the sublattices corresponding
to even and odd values of $j$ are separately conserved; this means
that $\langle c_j^\da c_{j\pm 1} \rangle = 0$ in any state independently
of the mean field approximation].
The final form of the mean field Hamiltonian is given by
\begin{eqnarray} H_{\text{MF}} &=&\sum_{j} ~[\alpha(c_{j}^{\da}c_{j+2}+
c_{j+2}^{\da}c_{j}) \non \\
&& ~~~~~~~+i\beta(c_{j}^{\da}c_{j+2}-c_{j+2}^{\da}c_{j}) -\mu' n_{j}], \non \\
\alpha &=& \cos(\phi/2), \non \\
\beta &=& 2\eta\sin(\phi/2). \label{hmf1} \end{eqnarray}
Transforming to momentum space, we find that the dispersion for the mean
field Hamiltonian in Fourier space is given by
\begin{equation} E_k =2\alpha\cos(2k)-2\beta\sin(2k)-\mu'. \label{disp}
\end{equation}

\begin{figure}[htb]
\centering
\includegraphics[width=0.46\textwidth]{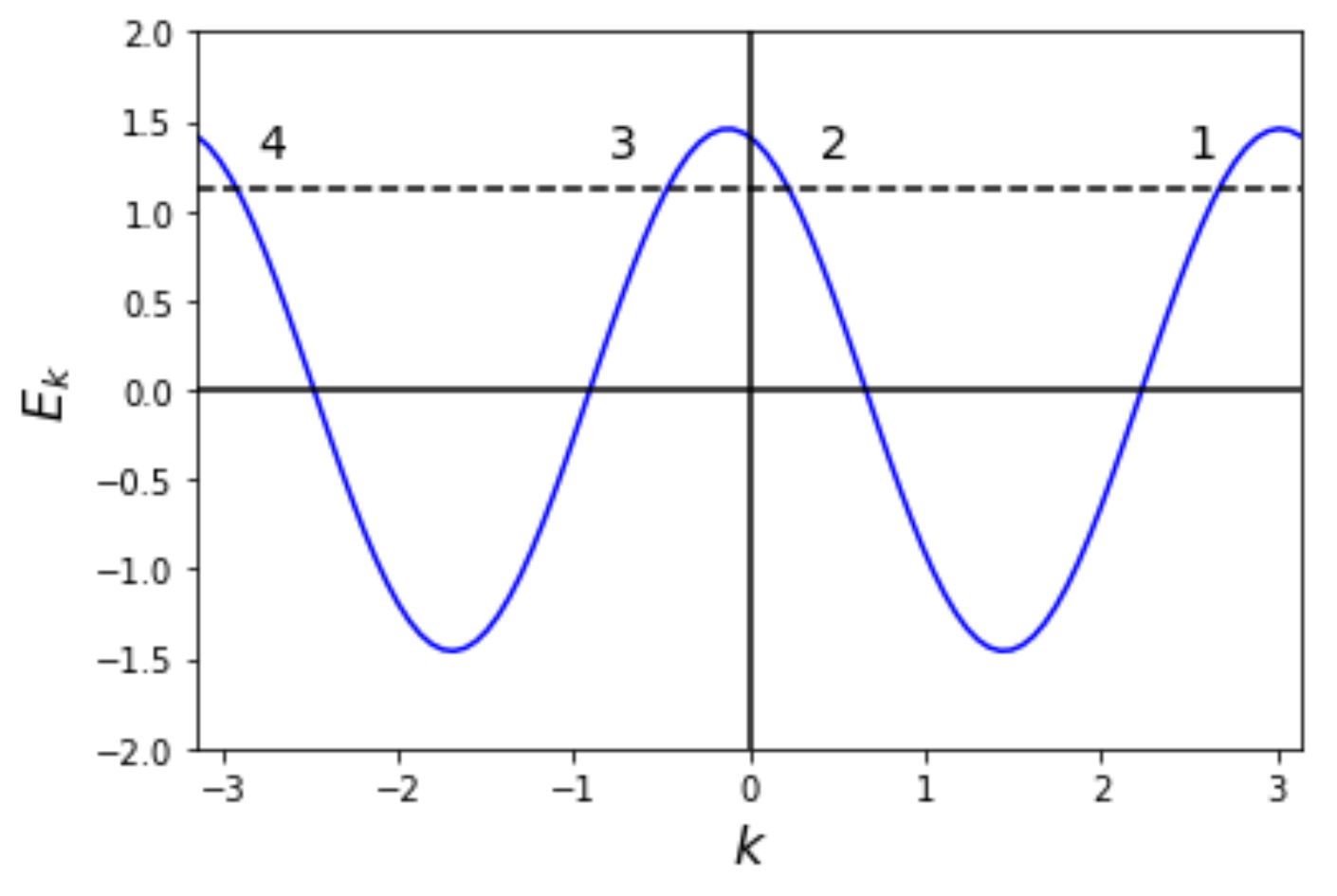}
\caption{Plot of the energy-momentum dispersion for $\phi = \pi/2$ and $\eta 
= 0.25$. The four Fermi points are labeled 1, 2, 3 and 4 from right to left, 
and the dotted line shows the value of $\mu'$.} \label{sa1fig02} \end{figure}

We now see that the mean field Hamiltonian in Eq.~\eqref{hmf1} again has four 
Fermi points for any value of $\phi$, provided that $\mu'$ lies in the range 
$-2 \sqrt{\alpha^2 + \beta^2} < \mu' < 2 \sqrt{\alpha^2 + \beta^2}$.
We denote the Fermi points by $k_{1}', k_{2}', k_{3}'$ and $k_{4}'$. 
The filling $1/2 + \eta$ leads to the following condition on the Fermi points,
\begin{equation} \frac{k_{1}'-k_{2}'+k_{3}'-k_{4}'}{2\pi}=\frac{1}{2}+\eta. 
\label{cond} \end{equation}
The Fermi points can be calculated using Eqs.~\eqref{disp} and \eqref{cond}.
We find that 
\begin{eqnarray}
k_{1}'&=&\frac{3\pi}{4}+\frac{\pi\eta}{2}-\frac{\theta}{2}, \non \\
k_{2}'&=&\frac{\pi}{4}-\frac{\pi\eta}{2}-\frac{\theta}{2}, \non \\
k_{3}'&=&-\frac{\pi}{4}+\frac{\pi\eta}{2}-\frac{\theta}{2}, \non \\
k_{4}'&=&-\frac{3\pi}{4}-\frac{\pi\eta}{2}-\frac{\theta}{2}, \label{k1234} 
\end{eqnarray}
where $\theta$ and $\mu'$ are given by
\begin{eqnarray}
\theta &=& \tan^{-1} (\beta/\alpha) ~=~ \tan^{-1}(2\eta\tan(\phi/2)), \non \\
\mu'&=& 2\sqrt{\alpha^2+\beta^2}\sin(\pi\eta). \label{thetamu} \end{eqnarray}
Note that all the Fermi points shift by the same amount given by $\theta /2$.
A plot of the dispersion in Eq.~\eqref{disp} along with the four Fermi points 
$k_{1}', k_{2}', k_{3}'$ and $k_{4}'$ from right to left is shown in 
Fig.~\ref{sa1fig02} for $\phi = \pi /2$ and $\eta = 0.25$.

We can check if any neighboring pair of Fermi points can cross each other as 
the filling $\eta$ and the statistical phase $\phi$ are varied. Let us 
discuss this crossing with respect to the Fermi point at $k'_{1}$. 
Equation~\eqref{k1234} implies that the momentum difference $|k'_1 - k'_2| =
\pi/2 + \pi \eta$ and $|k'_1 - k'_4| = \pi/2 - \pi \eta$ (to calculate the 
latter we have used the $2 \pi$ periodicity of $k$).
Since $\eta$ lies in the range $[-1/2, 1/2]$, we see that $k'_1$ and $k'_2$ 
can at most touch each other at $\eta=-1/2$ but not cross. Similarly $k'_1$ 
and $k'_4$ can at most touch at $\eta=1/2$ but not cross.

An important point to note from Eq.~\eqref{k1234} is that 
\beq k_{1}'-k_{3}' ~=~ k_{2}'-k_{4}' ~=~ \pi \label{kpi} \eeq
independently of $\eta$ and $\phi$. This is due to the fact that the 
Hamiltonian in Eq.~\eqref{ham4} remains invariant under the transformation 
\begin{eqnarray} c_{j}\:\rightarrow\:(-1)^{j}c_{j}, \non \\
c_{j}^{\da}\:\rightarrow\:(-1)^{j}c_{j}^{\da}. \label{transf2} \end{eqnarray}
Since $(-1)^j = e^{i j \pi}$, Eq.~\eqref{transf2} corresponds to shifting the 
momentum $k \to k + \pi$. As a result of this invariance, each Fermi point 
is accompanied by another Fermi point with a momentum difference $\pi$. [This 
symmetry of our model will be lost if we turn on a nearest-neighbor hopping]. 

The Fermi velocities $dE_k /dk$ at the four Fermi points can be calculated in 
terms of $\alpha$ and $\beta$ using Eqs.~\eqref{disp}, \eqref{k1234} and 
\eqref{thetamu},
\begin{eqnarray} v_{k_{1}'} &=& v_{k_{3}'} =v,\non\\
v_{k_{2}'} &=& v_{k_{4}'} =-v,\non\\
\text{where} ~~~v &=& 4\sqrt{\alpha^2+\beta^2}\cos(\pi\eta). 
\label{vel} \end{eqnarray}
Note that $v$ depends on both $\phi$ and $\eta$ and it vanishes at the 
limits $\eta = \pm 1/2$.
Equation~\eqref{vel} implies that fermions near both $k_{1}'$ and $k_{3}'$ are 
right-moving while those near $k_{2}'$ and $k_{4}'$ are left-moving.

\section{Bosonization}
\label{sec4}

We will now systematically develop a theory of fluctuations about the mean 
field theory using the technique of 
bosonization~\cite{gogolin,delft,rao,schulz,giam}. Bosonization involves
mapping interacting fermionic systems to systems of noninteracting
bosons in one dimension. It has been widely used to construct the low-energy 
theory of one-dimensional systems with two Fermi points, but our model 
requires us to use it for the case of four Fermi points. 

We begin by linearizing the spectrum near the four Fermi points in 
Eq.~\eqref{k1234} to obtain a low-energy effective description of the 
fluctuations. We write the fermionic annihilation operators as
\begin{eqnarray}
c_{j} &=& e^{ik'_1 j} ~\psi_1(j) ~+~ e^{ik'_2 j} ~\psi_2 (j) \non \\
&& +~ e^{ik'_3 j} ~\psi_3(j) ~+~ e^{ik'_4 j} ~\psi_4(j), 
\label{cj} \end{eqnarray}
where $k_{1}', ~k_{2}', ~k_{3}'$ and $k_{4}'$ are the four Fermi points, and 
$\psi_1, ~\psi_2, ~\psi_3$ and $\psi_4$ are the slowly varying
fields with momentum components lying near those points. We now rewrite the 
full Hamiltonian $:H_{\text{MF}} +H_{int}:$ (where $:~:$ denotes normal
ordering) in terms of these slowly varying fields. 
We consider the mean field part first. Using Eq.~\eqref{hmf1} and using the 
Taylor expansion $\psi(j+2)=\psi(j)+2\partial_{x}\psi(j) +$ higher order terms, 
the mean field Hamiltonian is given by
\begin{widetext}
\begin{eqnarray}
H&=&4i\int dx ~\big[(\alpha\sin(2k_{1}')+\beta\cos(2k_{1}')) ~\psi_1^{\da}
\partial_{x}\psi_1 ~+~ (\alpha\sin(2k_{2}')+\beta\cos(2k_{2}')) ~
\psi_2^{\da} \partial_{x}\psi_2 \non \\
&& ~~~~~~~~~~~~~~~+(\alpha\sin(2k_{3}')+\beta\cos(2k_{3}')) ~\psi_3^{\da}
\partial_{x} \psi_3 ~+~ (\alpha\sin(2k_{4}')+\beta\cos(2k_{4}')) ~
\psi_4^{\da} \partial_{x}\psi_4 \non \\
&& ~~~~~~~~~~~~~~~+(\alpha\cos(2k_{1}')-\beta\sin(2k_{1}')-\mu')~ \psi_{1}^{\da}
\psi_{1} ~+~ (\alpha\cos(2k_{2}')-\beta\sin(2k_{2}')-\mu') ~\psi_{2}^{\da}
\psi_{2} \non \\
&& ~~~~~~~~~~~~~~~+(\alpha\cos(2k_{3}')-\beta\sin(2k_{3}')-\mu') ~\psi_{3}^{\da}
\psi_{3} ~+~ (\alpha\cos(2k_{4}')-\beta\sin(2k_{4}')-\mu') ~\psi_{4}^{\da}
\psi_{4}\big]. \label{hmf2} \end{eqnarray}
\end{widetext}
This can be simplified by using Eqs.~\eqref{k1234}, \eqref{thetamu} and
\eqref{vel}. The Hamiltonian in Eq.~\eqref{hmf2} then reduces to
\begin{eqnarray}
H_{\text{MF}} &=& -i v \int dx ~[ \psi_1^{\da}\partial_{x}\psi_1 ~-~
\psi_2^{\da}\partial_{x} \psi_{2} \non \\
&& ~~~~~~~~~~~~~~~~~~+ \psi_{3}^{\da}\partial_{x}\psi_3 ~-~ \psi_4^{\da}
\partial_{x}\psi_4]. \label{hmf3} \end{eqnarray}
[Note that the coefficients of $\psi_{1}^{\da}\psi_{1}$, $\psi_{2}^{\da}
\psi_{2}$, $\psi_{3}^{\da}\psi_{3}$ and $\psi_{4}^{\da}\psi_{4}$ in 
Eq.~\eqref{hmf2} vanish due to Eqs.~(\ref{k1234}-\ref{thetamu})].

The Hamiltonian in Eq.~\eqref{hmf3} can be bosonized using the usual rules 
of bosonization. We use the following convention for the mapping between 
the fermionic and bosonic fields, $\psi_j (x)$ and $\phi_j (x)$,
\begin{eqnarray}
\psi_1 &=& F_{1}\frac{e^{-i2\sqrt{\pi}\phi_{1}}}{\sqrt{2\pi\alpha}}, \non \\
\psi_2 &=& F_{2}\frac{e^{i2\sqrt{\pi}\phi_{2}}}{\sqrt{2\pi\alpha}}, \non \\
\psi_3 &=& F_{3}\frac{e^{-i2\sqrt{\pi}\phi_{3}}}{\sqrt{2\pi\alpha}}, \non \\
\psi_4 &=& F_{4}\frac{e^{i2\sqrt{\pi}\phi_{4}}}{\sqrt{2\pi\alpha}},
\label{psiphi} \end{eqnarray}
where $\alpha$ is a short distance cut-off and $F_{j}$ ($j=1,2,3,4$) are 
the Klein factors which ensure the correct anticommutation relations between 
the fermionic operators. They have the following properties 
\begin{eqnarray}
F_{j}^{\da}F_{j}=F_{j}F_{j}^{\da}=1 ~~~{\rm for}~~~ j=1,2,3,4, \non \\
\{F_j,F_k\}=\{F_{j},F_{k}^{\da}\}=0, ~~~~{\rm for}~~~ j \ne k. \label{klein}
\end{eqnarray}
Using bosonization we find that Eq.~\eqref{hmf3} turns into
\begin{eqnarray}
H_{\text{MF}} &=& v \int dx ~[(\partial_{x}\phi_{1})^2 ~+~ (\partial_{x}
\phi_{2})^2 \non \\
&& ~~~~~~~~~~~~~~ + (\partial_{x}\phi_{3})^2 ~+~ (\partial_{x}\phi_{4})^2].
\label{hmfb1} \end{eqnarray}

We will now bosonize the interacting part of the Hamiltonian in 
Eq.~\eqref{ham10}, namely, $:H_{int}:$. 
The bosonization of this part is more complicated since there are many 
nonoscillatory terms possible with four fermionic fields. (We will ignore
all the rapidly oscillating terms since they average to zero when we integrate
over $x$). To examine these terms systematically, we will 
divide them into two categories, momentum conserving terms where the change 
in momentum $\Delta k=0$, and umklapp terms where $\Delta k=2\pi$.

We will consider the momentum conserving terms first. These terms can be 
further separated into two groups, interaction terms which involve only
one Fermi point (called diagonal density-density interactions below)
and interaction terms which involve two Fermi points (called off-diagonal
density-density interactions).
Using Eqs.~\eqref{cj} and \eqref{psiphi}, we get the following diagonal 
density-density interactions 
\begin{eqnarray} && H_{\text{diag}} ~=~ 4\sin(\phi/2) \non \\
&& \times \int dx \big[(\rho_{1}^2+\rho_{3}^2)\cos(\pi\eta-\theta) 
- (\rho_{2}^2+\rho_{4}^2)\cos(\pi\eta+\theta)\big], \non \\ 
\label{hdiag1} \end{eqnarray}
where $\rho_{j}=\psi_{j}^{\da} \psi_{j}$ ($j=1,2,3,4$), is the density 
operator in terms of fermionic fields. This Hamiltonian can be 
bosonized using the relation between the fermionic and bosonic fields
\begin{equation} \rho_{j}=-\frac{\partial_{x}\phi_{j}}{\sqrt{\pi}}, ~~~
\text{where}~~~ j=1,2,3,4. \label{rhophi} \end{equation}
Equation~\eqref{hdiag1} therefore has the bosonic form
\begin{eqnarray}
&& H_{\text{diag}} ~=~ \frac{4 \sin(\phi /2)}{\pi} \non \\
&& \times \int dx\big[\big((\partial_{x}\phi_{1})^2+(\partial_{x}\phi_{3})^2
\big) \cos(\pi\eta-\theta) \non \\
&& ~~~~~~~~~~~~~~-\big((\partial_{x}\phi_{2}\big)^2+\big(\partial_{x}
\phi_{4})^2 \big) \cos(\pi\eta+\theta)\big]. \label{hdiag2} \end{eqnarray}

Next we consider the off-diagonal density-density interactions which involve
two different fields. 
There are six terms possible which are given by
$\rho_{1}\rho_{2},~ \rho_{3}\rho_{4},~ \rho_{1}\rho_{3},~ \rho_{2}\rho_{3},
~\rho_{1}\rho_{3}$ and $\rho_{2}\rho_{4}$.
Using Eq.~\eqref{cj} and the fact that $k'_{1}+k'_{2}=\pi - \theta$ and 
$k'_{1}-k'_{2}=\pi/2+\pi\eta$, we find that the off-diagonal interactions are 
given by
\begin{widetext}
\begin{eqnarray}
H_{\text{off-diag}} &=& 8\sin(\phi/2)\sin(\theta)\int dx~ \big[(\sin(\pi\eta)+1)
(\rho_{1}\rho_{2}+\rho_{3}\rho_{4})+(\sin(\pi\eta)-1)(\rho_{1}\rho_{4}+\rho_{2}
\rho_{3})\big] \non \\
&& + ~16\sin(\phi/2)\int dx ~\big[\cos(\pi\eta-\theta)\rho_{1}\rho_{3}-
\cos(\pi\eta+\theta)\rho_{2}\rho_{4}\big]. \label{hoffd} \end{eqnarray}
\end{widetext}
Using Eq.~\eqref{rhophi}, the bosonized form of Eq.~\eqref{hoffd} is found 
to be
\begin{widetext}
\begin{eqnarray}
H_{\text{off-diag}}&=& \frac{8\sin(\phi/2)\sin(\theta)}{\pi} \int dx\big[
(\sin(\pi\eta)+1) \big((\partial_{x}\phi_{1})(\partial_{x}\phi_{2})+
(\partial_{x}\phi_{3}) (\partial_{x}\phi_{4})\big)\non \\
&& ~~~~~~~~~~~~~~~~~~~~~~~~~~~~~~~~~~~~~~~+(\sin(\pi\eta)-1)\big((\partial_{x}
\phi_{1})(\partial_{x}\phi_{4})+(\partial_{x}\phi_{2})(\partial_{x}\phi_{3})
\big)\big]\non\\ 
&& + ~\frac{16\sin(\phi/2)}{\pi} \int dx\big[\cos(\pi\eta-\theta)(\partial_{x}
\phi_{1})(\partial_{x}\phi_{3})-\cos(\pi\eta+\theta)(\partial_{x}\phi_{2})
(\partial_{x}\phi_{4})\big]. \end{eqnarray}
\end{widetext}

There is another kind of momentum conserving interaction possible in our 
system, which is a term involving all the four Fermi points. In the
fermionic language, these are given by
\begin{widetext}
\begin{eqnarray} H_{\text{four-Fermi-pt}} ~=~ -4i\sin(\phi/2)e^{-i\theta} \int 
dx ~\big[ \psi_{1}^{\da}\psi_{2}\psi_{4}^{\da}\psi_{3}+\psi_{2}^{\da}\psi_{1}
\psi_{3}^{\da}\psi_{4} ~+~ \sin(\pi\eta)\big(\psi_{1}^{\da}\psi_{3} 
\psi_{4}^{\da}\psi_{2}+\psi_{2}^{\da}\psi_{4}\psi_{3}^{\da}\psi_{1}\big)
\big] ~+~\text{H.c.} \label{4fermipt1} \end{eqnarray}
\end{widetext}
To convert this into the bosonic language, we use Eq.~\eqref{psiphi}.
We now get products of Klein factors. For instance, the first two terms in
Eq.~\eqref{4fermipt1} have the products $F_1^\da F_2 F_4^\da F_3$ and 
$F_2^\dag F_1 F_3^\da F_4$ respectively. Equations~\eqref{klein} imply that 
these two products commute with each other and can therefore be simultaneously 
diagonalized. Hence we can ignore the Klein factor products in the 
following~\cite{schulz}. We then find that in the bosonic language, 
Eq.~\eqref{4fermipt1} becomes
\begin{eqnarray} && H_{\text{four-Fermi-pt}} = \frac{16\sin(\phi/2)
\sin(\theta) (\sin(\pi\eta)-1)}{(2\pi\alpha)^2} \non \\
&& \times \int dx \cos [2\sqrt{\pi}(\phi_{1}+\phi_{2}-\phi_{3}-\phi_{4})].
\label{4fermipt2} \end{eqnarray}

We now discuss the various umklapp terms which appear. These describe 
scattering processes in which the momentum difference between the initial and 
final states is $2\pi$; these terms are allowed in a lattice system since
since a momentum transfer $\Delta k=2\pi$ is equivalent to $\Delta k=0$. In 
the bosonization of a system with two Fermi points (left and right-moving 
points denoted as $R$ and $L$ respectively), umklapp terms of the form 
$\psi_{R}^{\da}\psi_{R}^{\da}\psi_{L} \psi_{L}$ and its Hermitian conjugate 
are allowed at half-filling, i.e., with $k_F = \pi/2$, since $4 k_F = 2 \pi$.
In our system, however,there are two kinds of umklapp terms possible which 
are quite different from the umklapp terms which appear for the case of
two Fermi points. The first kind of umklapp term appears due to scattering 
between two Fermi points with the same chirality, both right- or both 
left-moving in same direction. This term is given by
\begin{widetext}
\begin{equation}
H_{\text{umklapp},1} ~=~ 2i\sin(\phi/2) \int dx \big[ e^{2ik_{3}'}
\psi_{1}^{\da} \psi_{3}\psi_{1}^{\da}\psi_{3} + e^{2ik_{1}'} \psi_{3}^{\da}
\psi_{1} \psi_{3}^{\da}\psi_{1} + e^{2ik_{4}'} \psi_{2}^{\da}\psi_{4}
\psi_{2}^{\da}\psi_{4} + e^{2ik_{2}'} \psi_{4}^{\da}\psi_{2}\psi_{4}^{\da}
\psi_{2} \big] ~+~ \text{H.c.} \label{umklapp1} \end{equation}
\end{widetext}
(Each term in this equation changes the momentum by $\pm 2 \pi$).
When we bosonize this, we again find that the products of Klein factors
(such as $(F_1^\da)^2 (F_3)^2$, $(F_3^\da)^2 (F_1)^2$, etc) all commute
with each other as well as with the products of Klein factors which appear
in Eq.~\eqref{4fermipt2}. Hence we can ignore all these product terms. We then
find that in the bosonic language, Eq.~\eqref{umklapp1} becomes
\begin{eqnarray} && H_{\text{umklapp},1} ~=~ - \frac{8\sin(\phi/2)\sin
(\theta)}{(2\pi \alpha)^2} \non \\
&& \times \int dx \big[\cos(4\sqrt{\pi}(\phi_{1}-\phi_{3}))+\cos(4\sqrt{\pi}
(\phi_{2}-\phi_{4}))\big]. \label{umklapp2} \non \\
\end{eqnarray}
The second kind of umklapp terms appears due to interaction among four 
Fermi points. This interaction is given by
\begin{widetext}
\begin{equation} H_{\text{umklapp},2}=2i\sin(\phi/2)\int dx\big[\psi_{1}^{\da}
\psi_{2}^{\da}\psi_{3}\psi_{4}(e^{-i2k_{1}'}+e^{-i2k_{2}'})+\psi_{3}^{\da}
\psi_{4}^{\da}\psi_{1}\psi_{2}(e^{-2ik_{3}'}+e^{-2ik_{4}'})\big]+
\text{H.c.} \label{umklapp3} \end{equation}
\end{widetext}
(Each term in Eq.~\eqref{umklapp3} changes the momentum by $\pm 2 \pi$, unlike 
the terms in Eq.~\eqref{4fermipt1} which conserve momentum). Upon bosonizing,
we again find that the products of Klein factors for the different terms
commute with each other and with all the products which appeared above; hence 
we ignore all these products. In the bosonic language, Eq.~\eqref{umklapp3} 
then becomes 
\begin{eqnarray} && H_{\text{umklapp},2} =\frac{16\sin(\phi/2)\sin(\theta)
\sin(\pi\eta)}{(2\pi\alpha)^2} \non \\
&& \times \int dx \cos [2\sqrt{\pi}(\phi_{1}-\phi_{2}-\phi_{3}+\phi_{4})].
\label{umklapp4} \end{eqnarray}
Note that in our system, umklapp terms appear at any filling due to 
Eq.~\eqref{kpi}, unlike systems with two Fermi points where umklapp terms 
appear only at half-filling where $4k_F = 2 \pi$.
 
The total Hamiltonian of our model is now given by
\begin{eqnarray} H_{\text{total}} &=&H_{\text{diag}} ~+~ H_{\text{off-diag}} ~
+~ H_{\text{four-Fermi-point}} \non \\
&& + ~H_{\text{umklapp},1} ~+~ H_{\text{umklapp},2}. 
\label{ham11} \end{eqnarray}
Next, we use the Bogoliubov transformation to diagonalize the quadratic part 
of the Hamiltonian given by $H_{\text{diag}}+H_{\text{off-diag}}$. 
It is convenient to write the quadratic part of the Hamiltonian in a matrix 
form. To do so, we choose the basis of bosonic fields
\begin{equation} \Phi= \begin{pmatrix} \phi_{1} &\phi_{2}& \phi_{3} &\phi_{4}
\end{pmatrix}^{T}, \end{equation}
where the subscript $T$ denotes the transpose of the row, so that $\Phi$ is
a column. In this basis, the quadratic part of Hamiltonian has the form
\begin{equation} H_{\text{quad}} ~=~ \int dx ~\pa_x \Phi^{T}M \pa_x \Phi, 
\label{mat1} \end{equation}
where $M$ is $4\times4$ matrix. We now define
\begin{eqnarray}
\alpha_{1}&=& v ~+~ \frac{4}{\pi}\sin(\phi/2)\cos(\pi\eta-\theta), \non \\
\alpha_{2}&=& v ~-~ \frac{4}{\pi}\sin(\phi/2)\cos(\pi\eta+\theta), \non \\
\alpha_{3}&=& \frac{4}{\pi}\sin(\phi/2)\sin(\theta)[\sin(\pi\eta)+1], \non \\
\alpha_{4}&=& \frac{4}{\pi}\sin(\phi/2)\sin(\theta)[\sin(\pi\eta)-1], \non \\
\alpha_{5}&=& \frac{8}{\pi}\sin(\phi/2)\cos(\pi\eta-\theta), \non \\
\alpha_{6}&=& -\frac{8}{\pi}\sin(\phi/2)\cos(\pi\eta+\theta). \label{alphai}
\end{eqnarray}
In terms of these parameters, the matrix $M$ in Eq.~\eqref{mat1} is given by
\begin{equation} M= \begin{pmatrix}
\alpha_{1} & \alpha_{3}&\alpha_{5}&\alpha_{4}\\
\alpha_{3}&\alpha_{2}&\alpha_{4}&\alpha_{6}\\
\alpha_{5}&\alpha_{4}&\alpha_{1}&\alpha_{3}\\
\alpha_{4}&\alpha_{6}&\alpha_{3}&\alpha_{2}
\end{pmatrix}. \end{equation}

We will now choose the following linear combinations of the bosonic fields,
\begin{eqnarray}
\phi_{R1}&=&\frac{\phi_{1}+\phi_{3}}{\sqrt{2}}, \non \\
\phi_{R2}&=&\frac{\phi_{1}-\phi_{3}}{\sqrt{2}}, \non \\
\phi_{L1}&=&\frac{\phi_{2}+\phi_{4}}{\sqrt{2}}, \non \\
\phi_{L2}&=&\frac{\phi_{2}-\phi_{4}}{\sqrt{2}}.
\label{new} \end{eqnarray}
In terms of creation and annihilation operators, the fields in Eq.~\eqref{new} 
are given by
\begin{eqnarray}
\phi_{Ri}&=&\chi_{Ri}+\chi_{Ri}^{\da}-\frac{\sqrt{\pi}x}{L}\hat{N}_{Ri}, ~~~
i\:=1,2, \non \\
\phi_{Li}&=&\chi_{Li}+\chi_{Li}^{\da}-\frac{\sqrt{\pi}x}{L}\hat{N}_{Li}, ~~~
i\:=1,2, \non \\
\chi_{Ri}&=&\frac{i}{2\sqrt{\pi}}\sum_{q>0}\frac{1}{\sqrt{n_{q}}}b_{Ri,q}
e^{iqx-\alpha q/2}, \non \\
\chi_{Ri}^{\da}&=&-\frac{i}{2\sqrt{\pi}}\sum_{q>0}\frac{1}{\sqrt{n_{q}}}
b_{Ri,q}^{\da} e^{-iqx-\alpha q/2}, \non \\
\chi_{Li}&=&-\frac{i}{2\sqrt{\pi}}\sum_{q>0}\frac{1}{\sqrt{n_{q}}}b_{Ri,q}
e^{-iqx-\alpha q/2}, \non \\
\chi_{Li}^{\da}&=&-\frac{i}{2\sqrt{\pi}}\sum_{q>0}\frac{1}{\sqrt{n_{q}}}
b_{Ri,q}^{\da} e^{iqx-\alpha q/2}.
\end{eqnarray}
where $L$ denotes the length of system (we are assuming periodic boundary
conditions so that a momentum $q$ can be defined as $q = 2 \pi n_q /L$, where
$n_q$ is an integer). We now define a new basis 
\begin{equation} \Phi'= \begin{pmatrix} 
\phi_{R1} &\phi_{L1}& \phi_{R2} &\phi_{L2} \end{pmatrix}^{T}. \end{equation}
In this basis the quadratic part of the Hamiltonian takes the form
\begin{equation} H_{\text{quad}}=\int dx ~\pa_x \Phi'^{T} M' \pa_x \Phi',
\label{ham12} \end{equation}
where $M'$ is given by
\begin{eqnarray} M'= \left( \begin{array}{cccc}
\alpha_{1}+\alpha_{5} & \alpha_{3}+\alpha_{4}&0&0 \\
\alpha_{3}+\alpha_{4}&\alpha_{2}+\alpha_{6}&0&0 \\
0&0&\alpha_{1}-\alpha_{5}&\alpha_{3}-\alpha_{4} \\
0&0&\alpha_{3}-\alpha_{4}&\alpha_{2}-\alpha_{6} \end{array} \right).
\label{mat2} \end{eqnarray}
We see that $M'$ has a block diagonal form. We will now diagonalize each 
block separately using the Bogoliubov transformation.

A remarkable point to note at this stage is that $\al_3 = \al_4 = 0$ if 
either the statistical phase $\phi = 0$ or $\eta = 0$ (i.e., half-filling). 
The matrix $M'$ is then diagonal. We therefore have a nontrivial interacting 
theory only if both $\phi$ and $\eta$ are nonzero (i.e., away from 
half-filling).

\subsection{Diagonalization of Hamiltonian and Luttinger parameters}
\label{sec4a}

We begin our analysis by considering the upper block of the Hamiltonian 
in Eq.~\eqref{ham12}. This can be written as 
\begin{equation} H_{\text{quad},1}=\int dx ~\pa_x \Phi_{1}^{'T} M'_{1}
\pa_x \Phi'_{1}, \label{ham13} \end{equation}
where $\Phi_{1}'$ is given by
\begin{equation} \Phi'_{1}= \begin{pmatrix} 
\phi_{R1} &\phi_{L1} \end{pmatrix}^{T}, \end{equation}
and 
\begin{eqnarray} M_{1}'= \begin{pmatrix}
\alpha_{1}+\alpha_{5} & \alpha_{3}+\alpha_{4} \\
\alpha_{3}+\alpha_{4}&\alpha_{2}+\alpha_{6} \end{pmatrix}. 
\label{m1} \end{eqnarray}

Next we define
\begin{eqnarray} v_{1}^{(1)}&=&\alpha_{1}+\alpha_{5}, \non \\
v_{2}^{(1)}&=&\alpha_{2}+\alpha_{6}, \non \\
\lambda^{(1)}&=&\alpha_{3}+\alpha_{4}. \end{eqnarray}
We then find that in momentum space, the Hamiltonian in Eq.~\eqref{ham13} 
is given, up to a constant, by
\begin{eqnarray}
H_{\text{quad},1} &=&\sum_{q>0} ~q\big[ v_{1}^{(1)}b_{q,R1}^{\da}b_{q,R1}+
v_{2}^{(1)}b_{q,L1}^{\da}b_{q,L1} \non \\
&& ~~~~~~~~~~+\lambda^{(1)}\big(b_{q,R1}^{\da}b_{q,L1}^{\da}+b_{q,L1}
b_{q,R1} \big) \big], \label{ham14} \end{eqnarray}
where $b_{q,R1}$ and $b_{q,L1}$ are bosonic operators obeying the usual 
commutation relations.
This Hamiltonian can be diagonalized using the Bogoliubov transformation 
\begin{eqnarray}
\Tilde{b}_{q,R1} &=&\frac{b_{q,R1}+\gamma_{1} {b}^\da_{q,L1}}{\sqrt{1-
\gamma_{1}^2}},\non\\
\Tilde{b}_{q,L1} &=&\frac{b_{q,L1}+\gamma_{1}b^\da_{q,R1}}{\sqrt{1-
\gamma_{1}^2}}, \non\\ 
\gamma_{1}&=&\frac{1-K_{1}}{1+K_{1}}.
\end{eqnarray}
The old and new $\phi$ fields in this block are related as
\begin{eqnarray}
\phi_{R1} &=&\frac{(1+K_1)\Tilde{\phi}_{R1}-(1-K_{1})\Tilde{\phi}_{L1}}{2
\sqrt{K_1}},\non\\
\phi_{L1} &=&\frac{(1+K_1)\Tilde{\phi}_{L1}-(1-K_{1})\Tilde{\phi}_{R1}}{2
\sqrt{K_1}}. \label{oldnew1} \end{eqnarray}
The Hamiltonian now takes the diagonalized form
\begin{eqnarray}
H_{\text{quad},1} &=& \sum_{q>0} ~q\big[\big\{v_{F}^{(1)}+\frac{v_{1}^{(1)}-
v_{2}^{(1)}}{2} \big\}\Tilde{b}_{q,R1}^{\da}\Tilde{b}_{q,R1} \non \\
&& ~~~~~~~+ \big\{v_{F}^{(1)}-\frac{v_{1}^{(1)}-v_{2}^{(1)}}{2}\big\}
\Tilde{b}_{q,L1}^{\da}\Tilde{b}_{q,L1}\big], \label{ham15} \end{eqnarray}
up to a constant, and $v_{F}^{(1)}$ and $K_{1}$ are given by
\begin{widetext}
\begin{eqnarray}
v_{F}^{(1)}&=&\frac{v_{1}^{(1)}+v_{2}^{(1)}}{2}\sqrt{1-\left(\frac{2
\lambda^{(1)}}{v_{1}^{(1)}+v_{2}^{(1)}}\right)^2},\non\\
&=&\frac{1}{2}\sqrt{\left(2v+\frac{24}{\pi}\sin(\phi/2)\sin(\pi\eta)\sin\theta
\right)^2 ~-~ \left(\frac{16}{\pi} \sin(\phi/2)\sin(\pi\eta)\right)^2},\non\\
K_{1}&=&\sqrt{\frac{v_{1}^{(1)}+v_{2}^{(1)}-2\lambda^{(1)}}{v_{1}^{(1)}+
v_{2}^{(1)}+2\lambda^{(1)}}} ~=~ \sqrt{\frac{2v+\frac{8}{\pi}\sin(\phi/2)
\sin(\theta)\sin(\pi\eta)}{2v+\frac{40}{\pi}\sin(\phi/2)\sin(\theta)
\sin(\pi\eta)}}. \label{vK1} \end{eqnarray}
\end{widetext}
We note from Eq.~\eqref{vK1} that the right- and left-moving bosonic fields 
do not have equal velocities; this is because our model breaks parity
symmetry ($x \to - x$) when $\phi \ne 0$. We can also find the condition for 
the ground state to be well-defined; as shown in Appendix A, the condition
turns out to be $v_{1}^{(1)}v_{2}^{(1)} > (\lambda^{(1)})^{2}$.

We now consider the lower block of $M'$ in Eq.~\eqref{mat2} which leads to
the Hamiltonian
\begin{equation} H_{\text{quad},2} ~=~ \int dx ~\pa_x \Phi_{2}^{'T} M'_{2}
\pa_x \Phi'_{2}, \label{ham16} \end{equation}
where
\begin{equation} \Phi'_{2}= \begin{pmatrix} 
\phi_{R2} &\phi_{L2} \end{pmatrix}^{T}, \end{equation}
and $M_{2}'$ has the form 
\begin{equation} M_{2}'= \begin{pmatrix}
\alpha_{1}-\alpha_{5} & \alpha_{3}-\alpha_{4} \\
\alpha_{3}-\alpha_{4}&\alpha_{2}-\alpha_{6} \end{pmatrix}. 
\label{m2} \end{equation}
We now define 
\begin{eqnarray} v_{1}^{(2)} &=&\alpha_{1}-\alpha_{5}, \non \\
v_{2}^{(2)} &=&\alpha_{2}-\alpha_{6}, \non \\
\lambda^{(2)} &=&\alpha_{3}-\alpha_{4}. \end{eqnarray}
In momentum space, the Hamiltonian in Eq.~\eqref{ham16} takes the form
\begin{eqnarray} H_{\text{quad},2} &=& \sum_{q>0} ~q\big[v_{1}^{(2)}
b_{q,R2}^{\da}b_{q,R2} ~+~ v_{2}^{(2)}b_{q,L2}^{\da}b_{q,L2} \non \\
&& ~~~~~~~~+\lambda^{(2)}\big(b_{q,R2}^{\da}b_{q,L2}^{\da}+b_{q,L2}b_{q,R2}
\big)\big] \label{hamx} \end{eqnarray}
up to a constant, where $b_{q,R2}$ and $b_{q,L}$ are bosonic fields obeying
the usual commutation relations. We then do a Bogoliubov transformation to
obtain a Hamiltonian in a diagonal form
\begin{eqnarray} H_{\text{quad},2} &=& \sum_{q>0}q\big[\big\{v_{F}^{(2)}+
\frac{v_{1}^{(2)}- v_{2}^{(2)}}{2} \big\} \Tilde{b}_{q,R2}^{\da}
\Tilde{b}_{q,R2} \non \\
&& ~~~~~~~~~+\big\{v_{F}^{(2)}-\frac{v_{1}^{(2)}-v_{2}^{(2)}}{2}\big\}
\Tilde{b}_{q,L2}^{\da}\Tilde{b}_{q,L2}\big], \end{eqnarray}
up to a constant, where the Bogoliubov transformation is
\begin{eqnarray} \Tilde{b}_{q,R2}&=&\frac{b_{q,R2}+\gamma_{2} 
{b}^\da_{q,L2}}{\sqrt{1-\gamma_{2}^2}},\non\\
\Tilde{b}_{q,L2}&=&\frac{b_{q,L2}+\gamma_{2}b^\da_{q,R2}}{\sqrt{1-
\gamma_{2}^2}}, \non\\
\gamma_{2}&=&\frac{1-K_{2}}{1+K_{2}}. \end{eqnarray}
The old and new $\phi$ fields are related as
\begin{eqnarray} \phi_{R2}&=&\frac{(1+K_2)\Tilde{\phi}_{R2}-(1-K_{2})
\Tilde{\phi}_{L2}}{2 \sqrt{K_2}},\non\\
\phi_{L2}&=&\frac{(1+K_2)\Tilde{\phi}_{L2}-(1-K_{2})\Tilde{\phi}_{R2}}{2
\sqrt{K_2}}, \label{oldnew2} \end{eqnarray}
where
\begin{widetext}
\begin{eqnarray} v_{F}^{(2)}&=&\frac{v_{1}^{(2)}+v_{2}^{(2)}}{2}\sqrt{1-\left(
\frac{2\lambda^{(2)}}{v_{1}^{(2)}+v_{2}^{(2)}}\right)^2},\non\\
&=&\frac{1}{2}\sqrt{\left(2v-\frac{8}{\pi}\sin(\phi/2)\sin(\pi\eta)\sin\theta
\right)^2 ~-~ \left( \frac{16}{\pi} \sin(\phi/2)\sin(\pi\eta)\right)^2},\non\\
K_{2}&=&\sqrt{\frac{v_{1}^{(2)}+v_{2}^{(2)}-2\lambda^{(2)}}{v_{1}^{(2)}+
v_{2}^{(2)}+2\lambda^{(2)}}} ~=~ \sqrt{\frac{2v-\frac{8}{\pi}\sin(\phi/2)
\sin(\theta)\left(\sin(\pi\eta)+2\right)}{2v-\frac{8}{\pi}\sin(\phi/2)\sin
(\theta)\left(\sin(\pi\eta)-2\right)}}. \label{vK2} \end{eqnarray}
\end{widetext}
We again see that the right- and left-moving bosonic fields have unequal 
velocities. As before the condition for the ground state to be well-defined 
turns out to be $v_{1}^{(2)}v_{2}^{(2)} > (\lambda^{(2)})^2$.

\begin{figure}[htb]
\centering
\subfigure[]{\includegraphics[width=8.8cm]{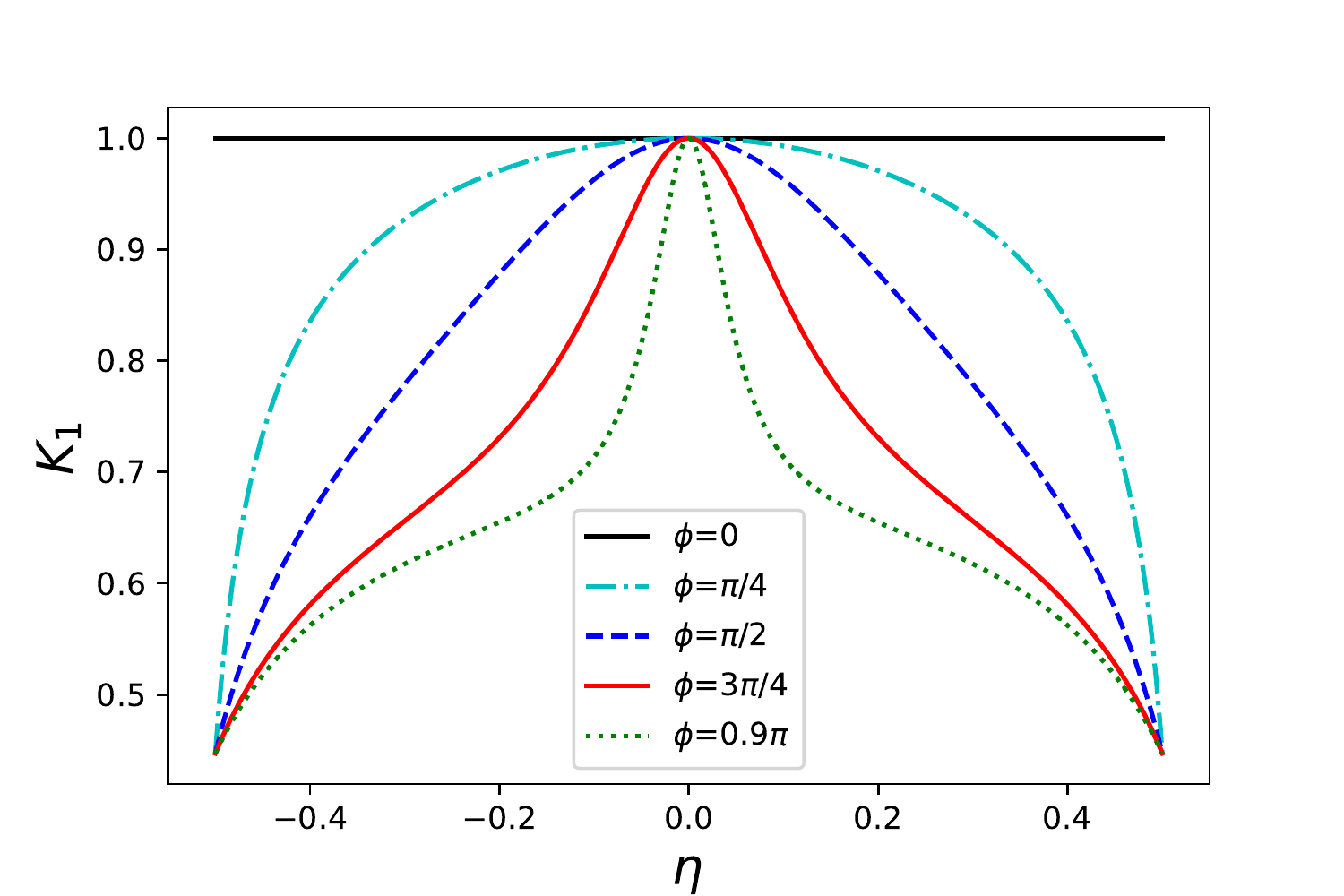}}
\subfigure[]{\includegraphics[width=8.8cm]{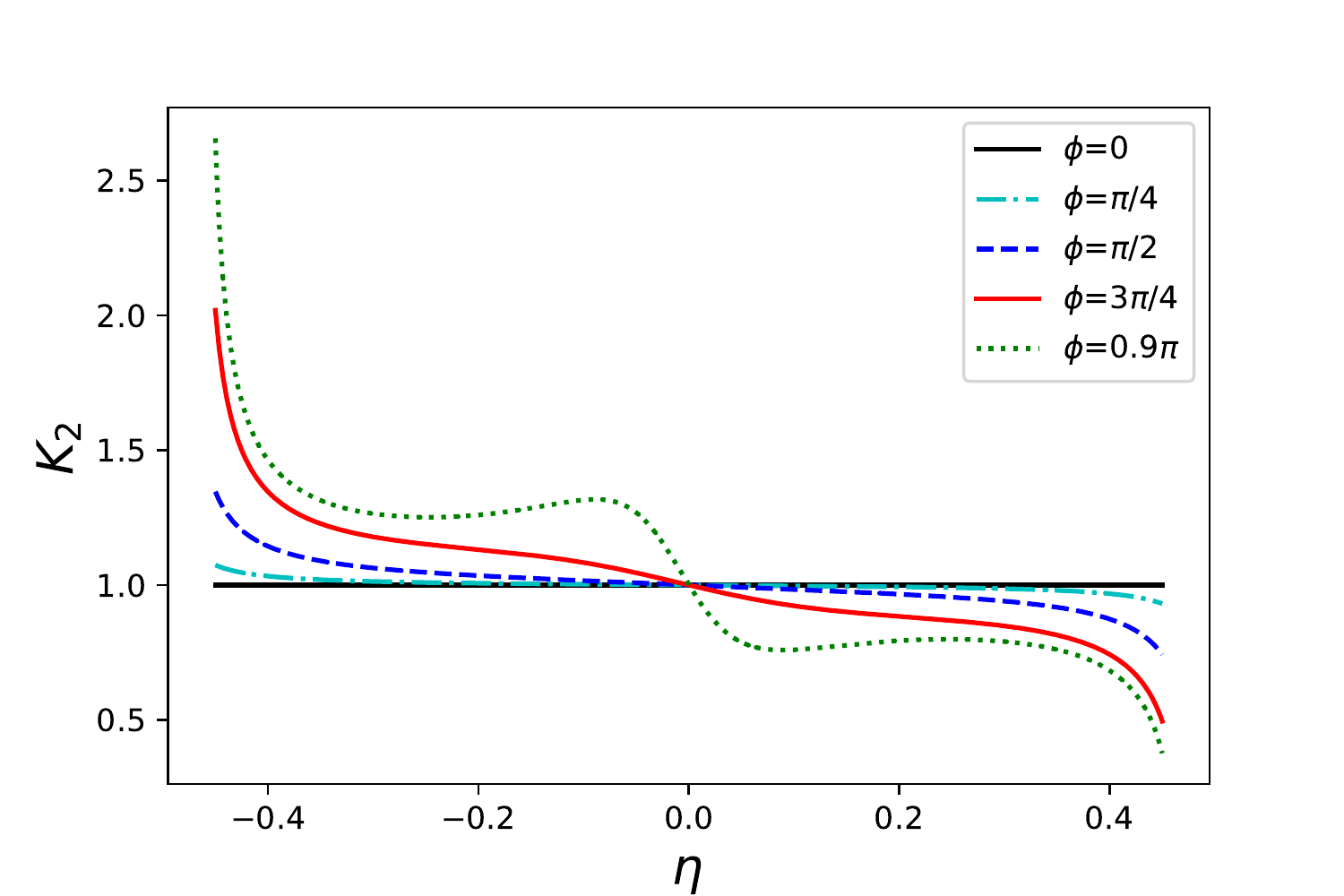}}
\caption{Luttinger parameters $K_1$ and $K_2$ vs $\eta$ for $\phi = 0, ~
\pi/4, ~\pi/2, ~3\pi/4$ and $0.9 \pi$. They satisfy the symmetries $K_1 (\phi,
-\eta)= K_1 (\phi,\eta)$ and $K_2 (\phi,-\eta)= 1/K_2 (\phi,\eta)$.} 
\label{sa1fig03} \end{figure} 

Figures~\ref{sa1fig03} (a-b) show $K_1$ and $K_2$ as functions of 
$\eta$ for various values of $\phi$. (We have not shown the values of $K_1$ 
and $K_2$ very close to $\eta = \pm 1/2$ since the analytical expressions in
Eqs.~\eqref{vK1} and \eqref{vK2} become singular in that limit. In particular, 
$v \to 0$ as $\eta \to \pm 1/2$). We see from the figures that $K_1$ remains 
unchanged while $K_2 \to 1/K_2$ if we flip $\eta \to - \eta$. The reason
for these symmetries is discussed in Sec.~\ref{sec4b}. We also see from 
Fig.~\ref{sa1fig03} that $K_1 \le 1$ for all values of $\eta$ and $\phi$,
while $K_2 \ge 1 ~(\le 1)$ for $\eta \le 0 ~(\ge 0)$ for all values of $\phi$.
This is in agreement with the expressions given in Eqs.~\eqref{vK1} and
\eqref{vK2}.

We conclude that the diagonalization of the quadratic parts of the bosonized
Hamiltonian gives two decoupled Tomonaga-Luttinger liquids. Since each of this
is described by a conformal field theory with central charge $c=1$, the
full system has $c=2$. 

\subsection{Implications of particle-hole and parity transformations for 
$K_1$ and $K_2$}
\label{sec4b}

In this section, we will discuss two kinds of transformations which leave
the Hamiltonian in Eq.~\eqref{ham4} invariant, and what these imply for the
Luttinger liquid parameters $K_1$ and $K_2$.

The first transformation that we will consider is particle-hole transformation,
keeping the statistical phase $\phi$ unchanged. We find that transforming 
\beq c_j ~\to ~e^{-ij\pi /2} ~c_j^\da \label{ph} \eeq
flips the sign of $n_{j+1} - 1/2$. This leaves the hopping part of the 
Hamiltonian in Eq.~\eqref{ham4} invariant 
but flips the sign of the chemical potential term, thus transforming 
the filling from $1/2 + \eta$ to $1/2 - \eta$. Looking at 
Eq.~\eqref{cj}, we see that Eq.~\eqref{ph} transforms the Fermi momenta as
\beq k'_i ~\to~ - ~k'_i ~+~ \frac{\pi}{2}. \label{transf3} \eeq
Further, $\eta \to - \eta$ implies that $\theta \to - \theta$ in 
Eq.~\eqref{thetamu}. Equation~\eqref{transf3} then means that $k'_1$ and
$k'_3$ get interchanged, and $k'_2$ and $k'_4$ remain as they are. 
Equations~\eqref{cj} and \eqref{psiphi} then imply that $\phi_1$ and $\phi_3$ 
get interchanged, and $\phi_2$ and $\phi_4$ remain unchanged. Turning to the 
parameters $v$ and $\al_i$ in Eqs.~\eqref{vel} and \eqref{alphai}, we see
that $v$, $\al_1$, $\al_2$, $\al_5$ and $\al_6$ remain unchanged, and
$\al_3$ and $\al_4$ get interchanged. We then see that the matrix
$M'_1$ remains unchanged in Eq.~\eqref{m1} while in Eq.~\eqref{m2}, the
off-diagonal terms flip sign. Equation~\eqref{vK1} then means that $K_1$ 
remains unchanged while Eq.~\eqref{vK2} means that $K_2$ transforms
to $1/K_2$ under $\eta \to - \eta$.

The second transformation we look at is $\phi \to - \phi$ combined with 
parity, $j \to - j$ in Eq.~\eqref{ham4} or $x \to - x$ in the continuum,
keeping the filling ($\eta$) unchanged.
This leaves Eq.~\eqref{ham4} unchanged. Equation~\eqref{thetamu} implies that
$\theta \to - \theta$ while Eq.~\eqref{k1234} means that $k'_1$ and
$k'_4$ get interchanged as do $k'_2$ and $k'_3$. Equations~\eqref{cj} and 
\eqref{psiphi} then imply that $\phi_1 \leftrightarrow - \phi_4$ and
$\phi_2 \leftrightarrow - \phi_3$. Further, the parameters $v$, $\al_3$ and
$\al_4$ remain unchanged, and $\al_1$ and $\al_2$ get interchanged as
do $\al_5$ and $\al_6$. Equations~\eqref{m1} and \eqref{m2} then show that the 
diagonal entries of $M'_1$ get interchanged while the off-diagonal entries
do not change, and similarly for $M'_2$. This means that $K_1$ and $K_2$
remain unchanged under $\phi \to - \phi$ as we can see from Eqs.~\eqref{vK1}
and \eqref{vK2}.

\subsection{Scaling dimensions of the various four-fermion interaction terms}
\label{sec4c}

In Sec.~\ref{sec4a} we diagonalized the quadratic part of the Hamiltonian 
and found the relation between the old and new bosonic fields. We will now 
discuss the scaling dimension of the various terms involving cosines of the 
bosonic fields (arising from four-fermion interacting terms) with respect 
to the new vacuum obtained after the Bogoliubov transformations. 

In general, two-point correlation functions of exponentials of bosonic fields 
decay as power laws, 
\begin{widetext}
\begin{eqnarray}
\bra{ \Tilde{0}}Te^{i2\sqrt{\pi}\beta\Tilde{\phi}_{R1}}e^{-i2\sqrt{\pi}\beta\Tilde{\phi}_{R1}}\ket{\Tilde{0}}&\sim& \left(\frac{\alpha}{v_{R}^{(1)}t-x-i\alpha\text{sign}(t)}\right)^{\beta^2},\non\\
\bra{ \Tilde{0}}Te^{i2\sqrt{\pi}\beta\Tilde{\phi}_{R2}}e^{-i2\sqrt{\pi}\beta\Tilde{\phi}_{R2}}\ket{\Tilde{0}}&\sim& \left(\frac{\alpha}{v_{R}^{(2)}t-x-i\alpha\text{sign}(t)}\right)^{\beta^2},\non\\
\bra{ \Tilde{0}}Te^{i2\sqrt{\pi}\beta\Tilde{\phi}_{L1}}e^{-i2\sqrt{\pi}\beta\Tilde{\phi}_{L1}}\ket{\Tilde{0}}&\sim& \left(\frac{\alpha}{v_{L}^{(1)}t+x-i\alpha\text{sign}(t)}\right)^{\beta^2},\non\\
\bra{ \Tilde{0}}Te^{i2\sqrt{\pi}\beta\Tilde{\phi}_{L2}}e^{-i2\sqrt{\pi}\beta\Tilde{\phi}_{L2}}\ket{\Tilde{0}}&\sim& \left(\frac{\alpha}{v_{L}^{(2)}t+x-i\alpha\text{sign}(t)}\right)^{\beta^2},\non\\
{\rm where} ~~~~~v_{R}^{(1)}= v_{F}^{(1)}+\frac{v_{1}^{(1)}-
v_{2}^{(1)}}{2},~~~~~~~
v_{R}^{(2)} &=& v_{F}^{(2)}+\frac{v_{1}^{(2)}-v_{2}^{(2)}}{2},\non\\
v_{L}^{(1)} = v_{F}^{(1)}-\frac{v_{1}^{(1)}-v_{2}^{(1)}}{2},~~~~~~~ 
v_{L}^{(2)} &=& v_{F}^{(2)}-\frac{v_{1}^{(2)}-v_{2}^{(2)}}{2}. 
\label{corrfn} \end{eqnarray}
\end{widetext}
Here $\Tilde{\phi}_{Ri}$ and $\Tilde{\phi}_{Li}$ are the new fields obtained 
after the Bogoliubov transformation and $\ket{\Tilde{0}}$ denotes the new 
vacuum. 

We first consider the operator in Eq.~\eqref{4fermipt2} given by
\beq A_1 ~=~ \cos(2\sqrt{2\pi}(\phi_{R2}+\phi_{L2})). \label{a1} \eeq
Equation~\eqref{oldnew2} implies that
\begin{equation}
\phi_{R2}+\phi_{L2}=\sqrt{K_2}(\Tilde{\phi}_{R2}+\Tilde{\phi}_{L2}).
\end{equation}
We find the scaling dimension of the operator in Eq.~\eqref{a1} by 
calculating the correlation function $\bra{\Tilde{0}}Te^{i2\sqrt{2\pi}
(\phi_{R2}+\phi_{L2})}e^{-i2\sqrt{2\pi} (\phi_{R2}+\phi_{L2})}
\ket{\Tilde{0}}$. 
Using Eq.~\eqref{corrfn}, we find that
\begin{widetext}
\begin{eqnarray}
\bra{\Tilde{0}}Te^{i2\sqrt{2\pi}(\phi_{R2}+\phi_{L2})}e^{-i2\sqrt{2\pi} 
(\phi_{R2}+\phi_{L2})}\ket{\Tilde{0}}
&=& \bra{\Tilde{0}}Te^{i2\sqrt{2\pi K_2} (\Tilde{\phi}_{R2} +\Tilde{\phi}_{L2})}e^{-i2\sqrt{2\pi K_2}(\Tilde{\phi}_{R2}+\Tilde{\phi}_{L2})}\ket{\Tilde{0}},
\non\\
&&\sim ~\left(\frac{\alpha}{v_{R}^{(2)}t-x-i\alpha\text{sign}(t)}\right)^{2K_2}
\left(\frac{\alpha}{v_{L}^{(2)}t+x-i\alpha\text{sign}(t)}\right)^{2K_2}.
\end{eqnarray}
\end{widetext}
Setting $t=0$, we conclude that at large spatial separation the correlation 
function falls off as $(\alpha/x)^{4K_{2}}$. This means that the operator 
in Eq.~\eqref{a1} has scaling dimension $2K_{2}$. This term is relevant 
in the renormalization group (RG) sense
if the scaling dimension is less than 2 which requires $K_2<1$. 

Next, we find the scaling dimension of the umklapp terms from the appropriate 
correlation functions. The operators in the first umklapp term in 
Eq.~\eqref{umklapp2} are
\beq A_2 ~=~ \cos(4\sqrt{2\pi}\phi_{R2})+\cos(4\sqrt{2\pi}\phi_{L2}). 
\label{a2} \eeq
To find the scaling dimension of these operators,
we calculate the correlation functions $\bra{\Tilde{0}}Te^{i4\sqrt{2\pi}
\phi_{R2}}e^{-i4\sqrt{2\pi} \phi_{R2}}\ket{\Tilde{0}}$ 
and $\bra{\Tilde{0}}Te^{i4\sqrt{2\pi}\phi_{L2}}e^{-i4\sqrt{2\pi}
\phi_{L2}}\ket{\Tilde{0}}$ respectively. We find that
\begin{widetext}
\begin{eqnarray}
\bra{\Tilde{0}}Te^{i4\sqrt{2\pi}\phi_{R2}}e^{-i4\sqrt{2\pi} \phi_{R2}}
\ket{\Tilde{0}}
&\sim& \left(\frac{\alpha}{v_{R}^{(2)}t-x-i\alpha\text{sign}(t)}
\right)^{\frac{2(1+K_2)^2}{K_2}}\left(\frac{\alpha}{v_{L}^{(2)}t+x-i\alpha
\text{sign}(t)}\right)^{\frac{2(1-K_2)^2}{K_2}},\non\\
\bra{\Tilde{0}}Te^{i4\sqrt{2\pi}\phi_{L2}}e^{-i4\sqrt{2\pi} \phi_{L2}}
\ket{\Tilde{0}}
&\sim& \left(\frac{\alpha}{v_{L}^{(2)}t+x-i\alpha\text{sign}(t)}\right)^{
\frac{2(1+K_2)^2}{K_2}_2}\left(\frac{\alpha}{v_{R}^{(2)}t-x-i\alpha
\text{sign}(t)}\right)^{\frac{2(1-K_2)^2}{K_2}}. \end{eqnarray}
\end{widetext}
So both the operators in Eq.~\eqref{a2} have the scaling dimension 
$2\left(K_2 + \frac{1}{K_2} \right)$. Since this is equal to or larger than 
4 for all values of $K_2$, this term is always irrelevant in the RG sense. 

Similarly we can calculate the scaling dimension of the operator in the
second umklapp term in Eq.~\eqref{umklapp4} given by
\beq A_3 ~=~ \cos (2\sqrt{2\pi}(\phi_{R2}-\phi_{L2})) \label{a3}, \eeq
by looking at the correlation function 
$\bra{\Tilde{0}}Te^{i2\sqrt{2\pi}(\phi_{R2}-\phi_{L2})}e^{-i2\sqrt{2\pi} 
(\phi_{R2}-\phi_{L2})}\ket{\Tilde{0}}$. This is given by 
\begin{widetext}
\begin{eqnarray}
\bra{\Tilde{0}}Te^{i2\sqrt{2\pi}(\phi_{R2}-\phi_{L2})}e^{-i2\sqrt{2\pi}
(\phi_{R2}- \phi_{L2})}\ket{\Tilde{0}}&\sim& \left(\frac{\alpha}{v_{R}^{(2)}t
-x-i\alpha
\text{sign}(t)}\right)^{\frac{2}{K_2}}\left(\frac{\alpha}{v_{L}^{(2)}t+x-i
\alpha\text{sign}(t)}\right)^{\frac{2}{K_2}}. \end{eqnarray}
\end{widetext}
The scaling dimension of this umklapp term is therefore $2/K_2$. This term 
is relevant if the dimension is less than $2$ which requires $K_2 >1$.

Given the scaling dimensions of the three operators $A_1$, $A_2$ and 
$A_3$, we can use RG equations to 
examine the effect that they would have on the long-distance 
properties of the model. Note that all of them involve fields belonging to only
the second block given in Eq.~\eqref{ham16}. Following Eqs.~\eqref{4fermipt2},
\eqref{umklapp2} and \eqref{umklapp4}, we can write the contributions
of these operators to the Hamiltonian as
\bea \de H &=& \int dx ~[\lm_1 A_1 ~+~ \lm_2 A_2 ~+~ \lm_3 A_3], \non \\
\lm_1 &=& \frac{16\sin(\phi/2) \sin(\theta) (\sin(\pi\eta)-1)}{(2\pi\alpha)^2},
\non \\
\lm_2 &=& - \frac{8\sin(\phi/2)\sin(\theta)}{(2\pi \alpha)^2}, \non \\
\lm_3 &=& \frac{16\sin(\phi/2)\sin(\theta)\sin(\pi\eta)}{(2\pi\alpha)^2}.
\label{lmiai} \eea
Given the scaling dimensions $2K_2$, $2(K_2 + 1/K_2)$ and $2/K_2$ of
the operators $A_1$, $A_2$ and $A_3$, we find that the coefficients $\lm_i$ in
Eq.~\eqref{lmiai} effectively become functions of the length scale $L$ 
and satisfy the RG equations
\bea \frac{d\lm_1}{dl} &=& (2 ~-~ 2K_2) ~\lm_1, \non \\
\frac{d\lm_2}{dl} &=& (2 ~-~ 2K_2 ~-~ \frac{2}{K_2}) ~\lm_2, \non \\
\frac{d\lm_3}{dl} &=& (2 ~-~ \frac{2}{K_2}) ~\lm_3, \label{rg} \eea
to first order in the $\lm_i$'s, where $l = \ln (L/a)$ and $a$ is the 
lattice spacing. These equations have to be solved with the initial
values of the $\lm_i$'s at $l=0$ (i.e., $L=a$) given in Eq.~\eqref{lmiai}. 
Equation~\eqref{rg} implies that the operator $A_2$ is always
irrelevant, i.e., $\lm_2 \to 0$ as $L \to \infty$ for any value of $K_2$.
The operator $A_1$ is relevant if $K_2 < 1$, i.e., if $\eta > 0$, while the
operator $A_3$ is relevant if $K_2 > 1$, i.e., if $\eta < 0$.
Hence, depending on the sign of $\eta$, either $\lm_1 \to \infty$ and
$\lm_3 \to 0$ or vice versa, as $L \to \infty$. Correspondingly, one of the
operators, $A_1$ or $A_3$, would get pinned to its minimum value, and small 
oscillations around that pinned value would then describe excitations with
a gap~\cite{giam}.

If either $\phi$ or $\eta$ is close to zero (the latter means that we 
are close to half-filling), then we see from Eqs.~\eqref{vK2} and \eqref{lmiai} 
that $K_2$ is close to 1 and the $\lm_i$'s are close to zero. Then the RG 
equations in Eq.~\eqref{rg} imply that the relevant coupling will grow and 
become of order 1 only at enormous values of the length scale $L/a$. For 
instance, suppose that $K_2$ is less than but close to 1. Then the first 
equation in 
Eq.~\eqref{rg} along with the value of $\lm_1 (0)$ given in Eq.~\eqref{lmiai} 
would imply that $\lm (l) \sim 1$ will occur at a length scale of the order of
\beq \frac{L}{a} ~\sim~ \left( \frac{1}{\lm_1 (0)} \right)^{1/(2 -2 K_2)}, \eeq
which will be very large number if $\lm_1 (0)$ and $1 - K_2$ are small.

However, all the above statements about RG flows are based only on the first 
order RG equations in Eq.~\eqref{rg}. When the relevant coupling grows, one 
should consider second order terms and see if those can lead to a nontrivial 
but gapless fixed point. More accurately, one should consider all the three 
operators $A_1, ~A_2, ~A_3$, derive RG equations up to second order in 
$\lm_1, ~\lm_2, \lm_3$ and $K_2$, and then study what these equations imply 
about the fate of the second block at long 
distances~\cite{cardy,affleck,dutta}. 
We note, however, that there are no perturbations in the first block of 
Tomonaga-Luttinger liquids. Hence, this block is expected to remain gapless 
and be described by a $c=1$ conformal field theory.

\subsection{Scaling dimensions of charge density and superconducting order 
parameters}
\label{sec4d}

In this section we will calculate the two-point correlation functions of 
charge density 
and superconducting order parameters and thereby find their scaling dimensions.
We first discuss the charge density order parameter; this corresponds
to the oscillating part of the density $\rho = c^\da c$. In systems with two 
Fermi points, the charge density wave (CDW) order parameter has the form
\begin{equation} O_{\text{CDW}}=\psi_{R}^{\da}\psi_{L}, \end{equation}
where $\psi_{R}$ and $\psi_{L}$ are the right- and left-moving fermions.
In our model, the CDW order parameter is more complicated since we have 
four Fermi points which implies that $\rho = \sum_{lm} \psi^\da_l \psi_m
e^{i (k'_m - k'_l) x}$ has oscillating terms whenever $l \ne m$. The CDW 
order parameter is therefore given by a sum of six terms,
\begin{eqnarray}
O_{\text{CDW}} &=& O_{1} ~+~ O_{2} ~+~ O_{3} ~+~ O_{4} ~+~ O_{5} ~+~ O_{6}, 
\non \\
O_{1} &=& \psi_{1}^{\da}\psi_{2},~~~~~ O_{2}=\psi_{1}^{\da}\psi_{4}, \non \\
O_{3} &=& \psi_{3}^{\da}\psi_{2},~~~~~ O_{4}=\psi_{3}^{\da}\psi_{4}, \non \\
O_{5} &=& \psi_{1}^{\da}\psi_{3},~~~~~ O_{6}=\psi_{2}^{\da}\psi_{4}, 
\end{eqnarray}
where $\psi_{1},\psi_{3}$ are right-moving fermions and 
$\psi_{2}, \psi_{4}$ are left-moving fermions. We need to calculate six 
correlation functions to find the scaling dimensions of $O_{1},\cdots,O_{6}$.
We find that the correlation functions $\bra{\Tilde{0}}O_{1}^{\da}(x,t)O_{1}
(0,0)\ket{\Tilde{0}}$, $\cdots$, $\bra{\Tilde{0}}O_{6}^{\da}(x,t)O_{6}(0,0)
\ket{\Tilde{0}}$ are given by 
\begin{widetext}
\begin{eqnarray} && \bra{\Tilde{0}}TO_{1}^{\da}(x,t)O_{1}(0,0)\ket{\Tilde{0}} 
~\sim~ \bra{\Tilde{0}}TO_{4}^{\da}(x,t)O_{4}(0,0)\ket{\Tilde{0}} \non \\
&& \sim \left(\frac{\alpha}{v_{L}^{(1)}t+x-i\alpha\text{sign}(t)}\right)^{\frac{K_{1}}{2}} \left(\frac{\alpha}{v_{R}^{(1)}t-x-i\alpha\text{sign}(t)}\right)^{\frac{K_1}{2}} \left(\frac{\alpha}{v_{L}^{(2)}t+x-i\alpha\text{sign}(t)}\right)^{\frac{K_2}{2}} \left(\frac{\alpha}{v_{R}^{(2)}t-x-i\alpha\text{sign}(t)}\right)^{\frac{K_2}{2}},\non\\
\non \\
&& \bra{\Tilde{0}}TO_{2}^{\da}(x,t)O_{2}(0,0)\ket{\Tilde{0}} ~\sim~
\bra{\Tilde{0}}TO_{3}^{\da}(x,t)O_{3}(0,0)\ket{\Tilde{0}} \non \\
&& \sim \left(\frac{\alpha}{v_{L}^{(1)}t+x-i\alpha\text{sign}(t)}\right)^{\frac{K_{1}}{2}} \left(\frac{\alpha}{v_{R}^{(1)}t-x-i\alpha\text{sign}(t)}\right)^{\frac{K_1}{2}} \left(\frac{\alpha}{v_{L}^{(2)}t+x-i\alpha\text{sign}(t)}\right)^{\frac{1}{2K_2}} \left(\frac{\alpha}{v_{R}^{(2)}t-x-i\alpha\text{sign}(t)}
\right)^{\frac{1}{2K_2}}, \non \\
&& \bra{\Tilde{0}}TO_{5}^{\dagger}(x,t)O_{5}(0,0)\ket{\Tilde{0}} ~\sim~ 
\left(\frac{\alpha}{v_{L}^{(2)}t+x-i\alpha\text{sign}(t)}\right)^{\frac{(1-K_2)^2}{2K_2}}\left(\frac{\alpha}{v_{R}^{(2)}t-x-i\alpha\text{sign}(t)}\right)^{\frac{(1+K_2)^{2}}{2K_2}},\nonumber\\
&& \bra{\Tilde{0}}TO_{6}^{\dagger}(x,t)O_{6}(0,0)\ket{\Tilde{0}} ~\sim~ 
\left(\frac{\alpha}{v_{L}^{(2)}t+x-i\alpha\text{sign}(t)}\right)^{\frac{(1+K_2)^2}{2K_2}}\left(\frac{\alpha}{v_{R}^{(2)}t-x-i\alpha\text{sign}(t)}\right)^{\frac{(1-K_2)^{2}}{2K_2}}.
\end{eqnarray}
\end{widetext}
From the power law fall-offs at large spatial separations (setting $t=0$),
we see that 
$O_{1}$ and $O_{4}$ have scaling dimension $\frac{K_1}{2} + \frac{K_2}{2}$ and 
are relevant if $K_1 + K_2 < 4$, $O_{2}$ and $O_{3}$ have scaling dimension 
$\frac{K_1}{2} + \frac{1}{2K_2}$ and are relevant if $K_1 + \frac{1}{K_2} < 4$,
and $O_{5}$ and $O_{6}$ have scaling dimension $\frac{K_2}{2} 
+ \frac{1}{2K_2}$ and are relevant if $K_2 + \frac{1}{K_2} < 4$.

We now discuss the superconducting order parameters and their scaling 
dimensions. In our model there are six such terms whose sum is given by
\begin{eqnarray}
O_{\text{SC}}&=& O_{1}' ~+~ O_{2}' ~+~ O_{3}' ~+~ O_{4}' ~+~ O_{5}' ~+~ O_{6}',
\non \\
O_{1}'&=&\psi_{1}^{\da}\psi_{2}^{\da},~~~~~
O_{2}' ~=~ \psi_{1}^{\da}\psi_{4}^{\da}, \non \\
O_{3}'&=&\psi_{3}^{\da}\psi_{2}^{\da},~~~~~
O_{4}'~=~ \psi_{3}^{\da}\psi_{4}^{\da}, \non \\
O_{5}'&=&\psi_{1}^{\da}\psi_{3}^{\da},~~~~~
O_{6}'~=~ \psi_{2}^{\da}\psi_{4}^{\da}. \end{eqnarray}
Calculating the correlation functions of these order parameters similarly,
we find that
\begin{widetext}
\begin{eqnarray}
&& \bra{\Tilde{0}}TO_{1}'^{\da}(x,t)O_{1}'(0,0)\ket{\Tilde{0}} ~\sim~
\bra{\Tilde{0}}TO_{3}'^{\da}(x,t)O_{3}'(0,0)\ket{\Tilde{0}} \non \\
&& \sim \left(\frac{\alpha}{v_{L}^{(1)}t+x-i\alpha\text{sign}(t)}\right)^{\frac{1}{2K_1}}\left(\frac{\alpha}{v_{R}^{(1)}t-x-i\alpha\text{sign}(t)}\right)^{\frac{1}{2K_1}} \left(\frac{\alpha}{v_{L}^{(2)}t+x-i\alpha\text{sign}(t)}\right)^{\frac{1}{2K_2}}\left(\frac{\alpha}{v_{R}^{(2)}t-x-i\alpha\text{sign}(t)}\right)^{\frac{1}{2K_2}},\non\\
\non \\
&& \bra{\Tilde{0}}TO_{2}'^{\da}(x,t)O_{2}'(0,0)\ket{\Tilde{0}} ~\sim~
\bra{\Tilde{0}}TO_{4}'^{\da}(x,t)O_{4}'(0,0)\ket{\Tilde{0}} \non \\
&& \sim \left(\frac{\alpha}{v_{L}^{(1)}t+x-i\alpha\text{sign}(t)}\right)^{\frac{1}{2K_1}}\left(\frac{\alpha}{v_{R}^{(1)}t-x-i\alpha\text{sign}(t)}\right)^{\frac{1}{2K_1}} ~\left(\frac{\alpha}{v_{L}^{(2)}t+x-i\alpha\text{sign}(t)}\right)^{\frac{K_2}{2}}\left(\frac{\alpha}{v_{R}^{(2)}t-x-i\alpha\text{sign}(t)}\right)^{\frac{K_2}{2}}, \non \\
&& \bra{\Tilde{0}}TO_{5}'^{\dagger}(x,t)O_{5}'(0,0)\ket{\Tilde{0}} ~\sim~ 
\left(\frac{\alpha}{v_{L}^{(1)}t+x-i\alpha\text{sign}(t)}\right)^\frac{(1-K_1)^2}{2K_{1}}\left(\frac{\alpha}{v_{R}^{(1)}t-x-i\alpha\text{sign}(t)}\right)^{\frac{(1+K_1)^2}{2K_1}},\nonumber\\
&& \bra{\Tilde{0}}TO_{6}'^{\dagger}(x,t)O_{6}'(0,0)\ket{\Tilde{0}} ~\sim~
\left(\frac{\alpha}{v_{L}^{(1)}t+x-i\alpha\text{sign}(t)}\right)^\frac{(1+K_1)^2}{2K_{1}}\left(\frac{\alpha}{v_{R}^{(1)}t-x-i\alpha\text{sign}(t)}\right)^{\frac{(1-K_1)^2}{2K_1}}.
\end{eqnarray}
\end{widetext}
The power law fall-offs at large spatial separations imply that 
$O_{1}'$ and $O_{3}'$ have scaling dimension $\frac{1}{2K_1}+\frac{1}{2K_2}$ 
and are relevant if $\frac{1}{K_1}+\frac{1}{K_2}<4$, $O_{2}'$ and 
$O_{4}'$ have scaling dimension $\frac{1}{2K_1}+\frac{K_2}{2}$ 
and are relevant if $\frac{1}{K_1}+K_2<4$, and $O_{5}'$ and 
$O_{6}'$ have scaling dimension $\frac{K_1}{2} + \frac{1}{2K_1}$
and are relevant if $K_1 + \frac{1}{K_1} <4$.

\begin{figure}[htb]
\centering
\includegraphics[width=0.52\textwidth]{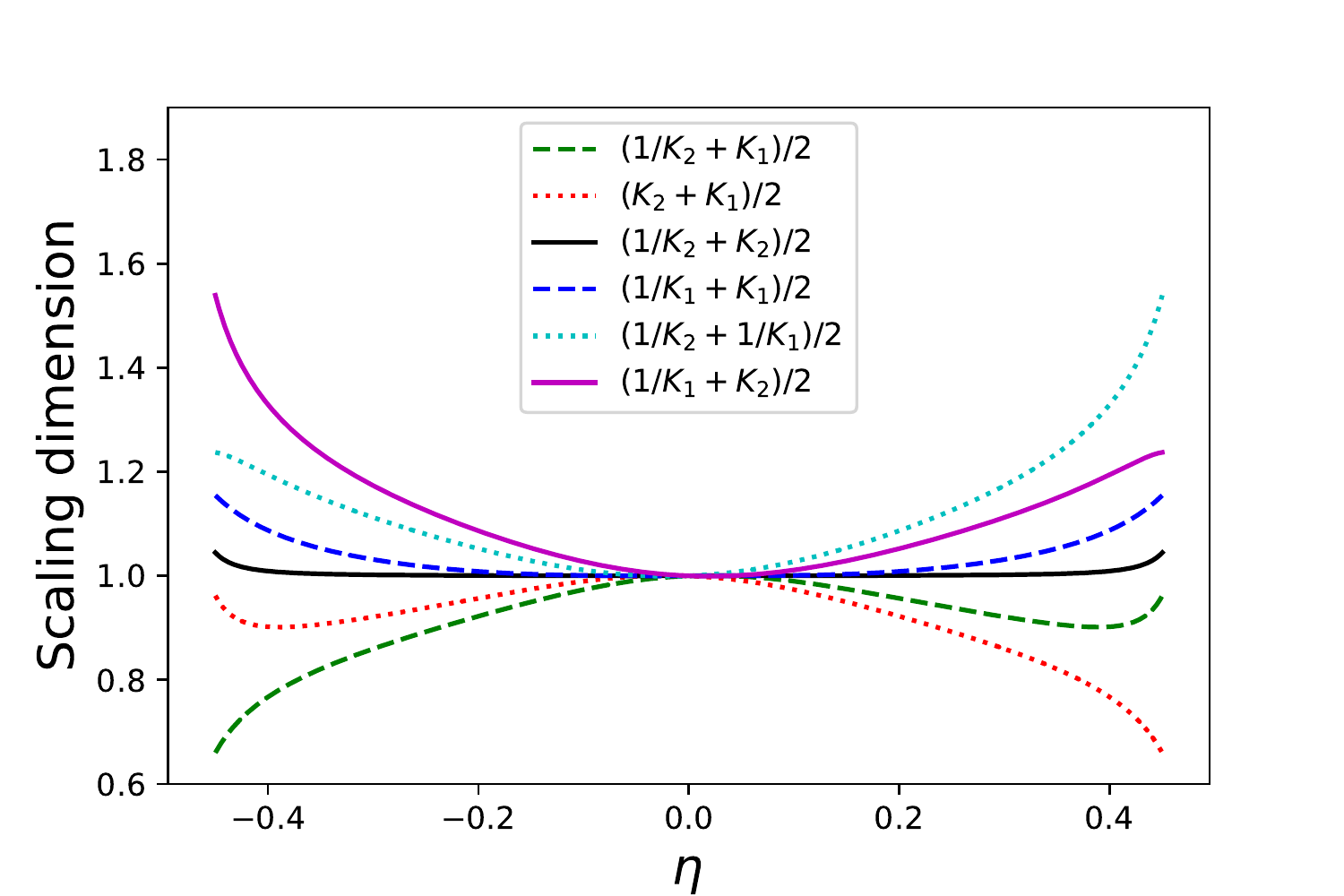}
\caption{Plots of scaling dimensions of the charge density and superconducting
order parameters vs $\eta$ for $\phi = \pi/2$. We see that the most
relevant operator is one of the charge density order parameters with scaling 
dimension $\frac{K_1}{2} + \frac{1}{2K_2}$ for $\eta < 0$ and $\frac{K_1}{2} 
+ \frac{K_2}{2}$ for $\eta > 0$.} \label{sa1fig04} \end{figure}

When both $K_1$ and $K_2$ are close to 1, all the six order parameters 
(three charge density and three superconducting) have scaling dimension close
to 1 and are therefore relevant. However, it is interesting to see which
of them has the smallest scaling dimension (i.e., the two-point correlation
function decays with the smallest power) and is thus the most 
relevant. In Fig.~\ref{sa1fig04}, we plot the six scaling dimensions as
functions of $\eta$ for $\phi = \pi/2$. We find that in this case, one
of the charge density order parameters has the smallest scaling dimension
for all values of $\eta$: for $\eta < 0$, the most relevant order parameter
is $\psi_1^\da \psi_4$ and $\psi_3^\da \psi_2$ with scaling dimension 
$\frac{K_1}{2} + \frac{1}{2K_2}$, while for $\eta > 0$, the most relevant 
order parameter is $\psi_1^\da \psi_2$ and $\psi_3^\da \psi_4$ with scaling 
dimension $\frac{K_1}{2} + \frac{K_2}{2}$.

\section{Two-particle bound states}
\label{sec5}

In this section, we will study what happens if the system has only
two particles, in particular, if there are two-particle bound states.
From Eq.~\eqref{ham4} it is clear that there are no interactions between
the two particles if they are on the same sublattice (i.e., both have
$j$ even or $j$ odd). We will therefore consider the case where one particle
is at site $n_1$ which is odd and the other particle is at site $n_2$ which
is even. We will define two-particle states as 
\beq |n_1,n_2 \rangle ~=~ c_{n_1}^\da c_{n_2}^\da |{\rm vacuum} \rangle , \eeq
regardless of whether $n_1 < n_2$ or $n_1 > n_2$. Next, we consider states
of the form
\beq |\psi \rangle ~=~ \sum_{n_1 ~{\rm odd}} \sum_{n_2 ~{\rm even}} 
\psi (n_1, n_2) ~|n_1,n_2 \rangle. \eeq
Since the center of mass is insensitive to interactions, we will
consider wave functions of the form
\beq \psi (n_1, n_2) ~=~ e^{iP (n_1 + n_2)/2} ~f(n_2 - n_1), \label{psi} \eeq
where $P$ is the center-of-mass momentum, and the relative coordinate wave 
function $f (n_2 - n_1)$ can depend on $P$. Since the wave function only 
changes by a minus sign if $P$ is shifted by $2\pi$ (since $n_1 + n_2$ is
an odd integer), we can take $P$ to lie in the range $[- \pi, \pi]$.

Given the Hamiltonian in Eq.~\eqref{ham4} (we will set $\mu =0$ in this
section), we find that the eigenvalue condition $H | \psi \rangle = E 
| \psi \rangle$ implies that the wave function $f (n)$ (where $n = n_2 - n_1$
is an odd integer) must satisfy
\bea && - 2 t_2 ~\cos (P ~-~ \frac{\phi}{2}) ~[f (n-2) ~+~ f(n+2)] \non \\
&=& E f(n) ~~{\rm for}~~ |n| ~\ge~ 3, \non \\
&& -2 t_2 ~[\cos (P ~-~ \frac{\phi}{2}) ~f(3) ~+~ \cos (P ~+~ 
\frac{\phi}{2}) ~f(-1)] \non \\
&=& E f (1), \non \\
&& -2 t_2 ~[\cos (P ~-~ \frac{\phi}{2}) ~f(-3) ~+~ \cos (P ~+~ 
\frac{\phi}{2}) ~f(1)] \non \\
&=& E f (-1). \label{fn} \eea
Equations~\eqref{fn} describe a particle moving on a lattice with only 
odd numbered 
sites, where the hopping amplitude between sites labeled $-1$ and $+1$ is $-2 
t_2 \cos (P+ \phi/2)$ and the hopping amplitude between all other neighboring
sites is $-2t_2 \cos (P - \phi /2)$. This system clearly has scattering states
for which 
\beq f(n_2 - n_1) ~=~ e^{i k (n_2 - n_1)/2} \eeq
for $|n_2 - n_1| \gg 1$, where $k$ lies in the range $[-\pi,\pi]$. 
The energy of such a state is
\beq E (P,k) ~=~ -4 t_2 \cos (P ~-~ \frac{\phi}{2}) ~\cos k, \label{epk} \eeq
which is simply the sum of the energies $E(k_1) = -2 t_2 \cos (2k_1 - \phi/2)$
and $E(k_2) = -2 t_2 \cos (2k_2 - \phi/2)$ of two particles 
moving independently on odd and even numbered sites with momenta 
\beq k_1 ~=~ \frac{P-k}{2} ~~~{\rm and}~~~ k_2 ~=~ \frac{P-k}{2} \label{k1k2}
\eeq
respectively. These form a band with energies going from $- |4 t_2 \cos (P
- \phi/2)|$ to $+ |4 t_2 \cos (P - \phi/2)|$.

We now examine if this relative coordinate also has bound states in addition
to the continuum of scattering states discussed above. The wave function
of such states must go to zero exponentially as $|n_2 - n_1| \to \infty$.
We therefore make the ansatz that the bound state wave function and energy are
given by
\bea f(n_2 - n_1) &=& e^{-\chi (n_2 - n_1)/2} ~~~{\rm for}~~~ n_2 - n_1 
\ge 1, \non \\
&=& \pm ~e^{\chi (n_2 - n_1)/2} ~~~{\rm for}~~~ n_2 - n_1 \le -1, \non \\
{\rm and}~~~ E (P,\chi) &=& -4 t_2 \cos (P ~-~ \frac{\phi}{2}) ~\cosh \chi,
\label{bound} \eea
where the real part of $\chi$ is positive. We then find that such bound states
exist if 
\beq | \frac{\cos (P ~+~ \frac{\phi}{2})}{\cos (P ~-~ \frac{\phi}{2})} | ~>~ 1,
\label{boundcond} \eeq
in which case $\chi$ is equal to either $r$ or $r + i \pi$, where 
$r$ is a positive real number in both cases, i.e., $\cosh \chi$ is $> 0$ or
$< 0$, and 
\beq e^r ~=~ | \frac{\cos (P ~+~ \frac{\phi}{2})}{\cos (P ~-~
\frac{\phi}{2})} |. \eeq
Further, if bound states exist, they appear in pairs with equal and opposite
energies given by
\beq E ~=~ \pm 4 t_2 \cos (P - \phi/2) \cosh r. \label{ebound} \eeq
The $\pm$ sign in the second line of Eq.~\eqref{bound} and the sign of $\cosh 
\chi$ depend on the signs of $\cos (P + \phi/2)$, $\cos (P - \phi/2)$ and $E$.

The condition in Eq.~\eqref{boundcond} implies that there are no bound states 
if $\phi = 0$ or $\pi$. If $0 < \phi < \pi$, Eq.~\eqref{boundcond} means 
that there are bound states if either $- \pi/2 < P < 0$ or $\pi/2 < P < \pi$.

\begin{widetext}
\begin{center} 
\begin{figure}[htb]
\subfigure[]{\includegraphics[width=5.9cm]{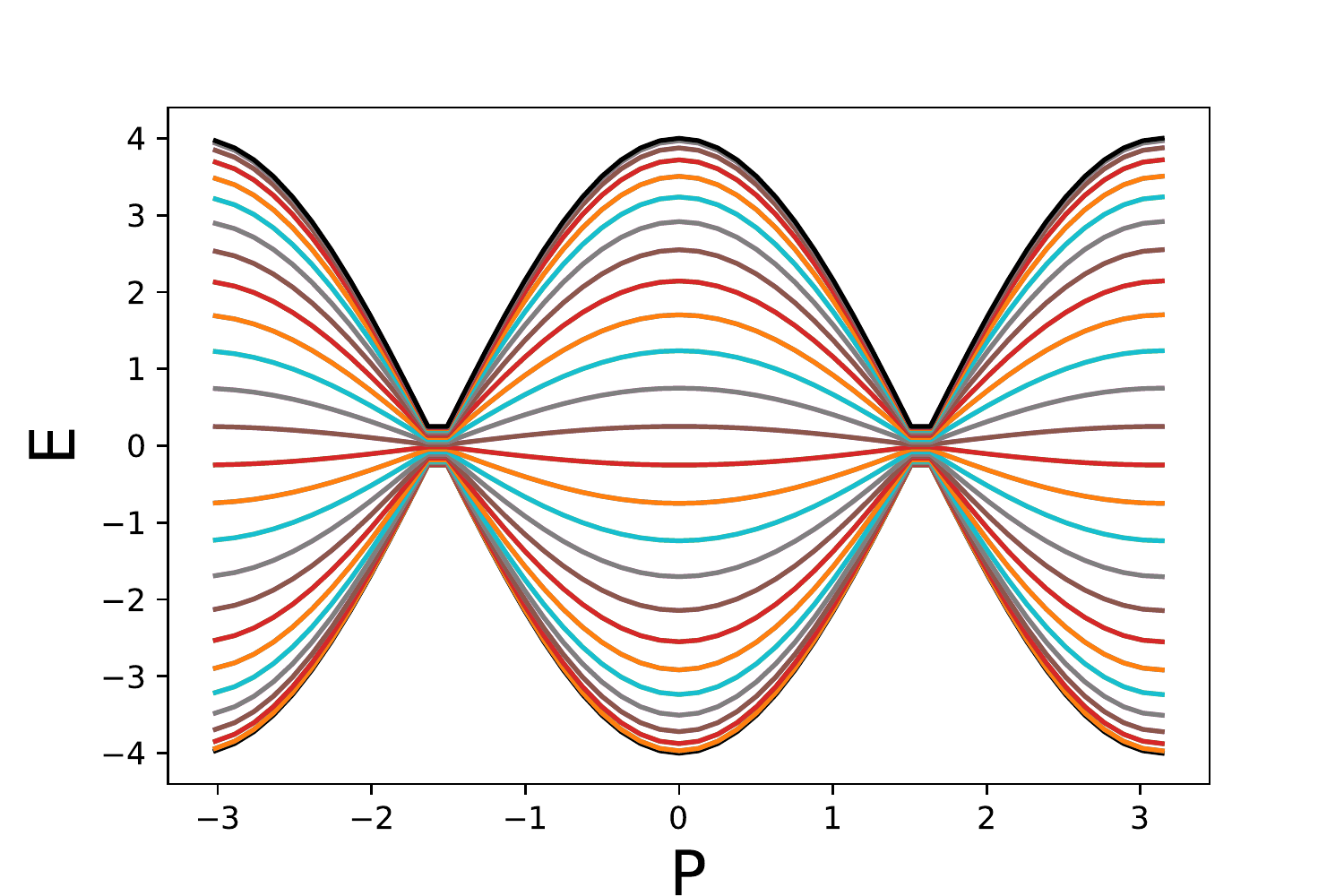}}
\subfigure[]{\includegraphics[width=5.9cm]{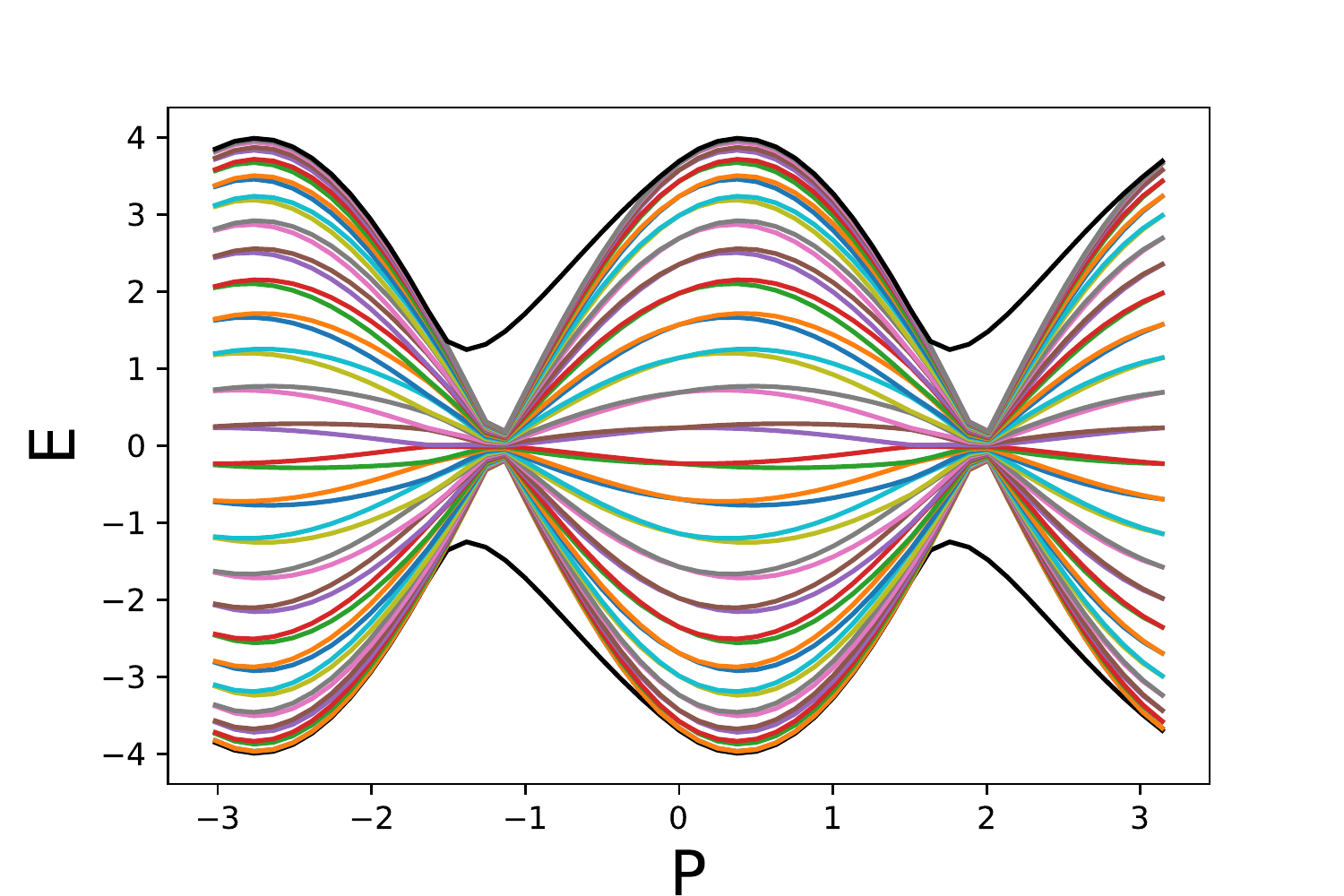}}
\subfigure[]{\includegraphics[width=5.9cm]{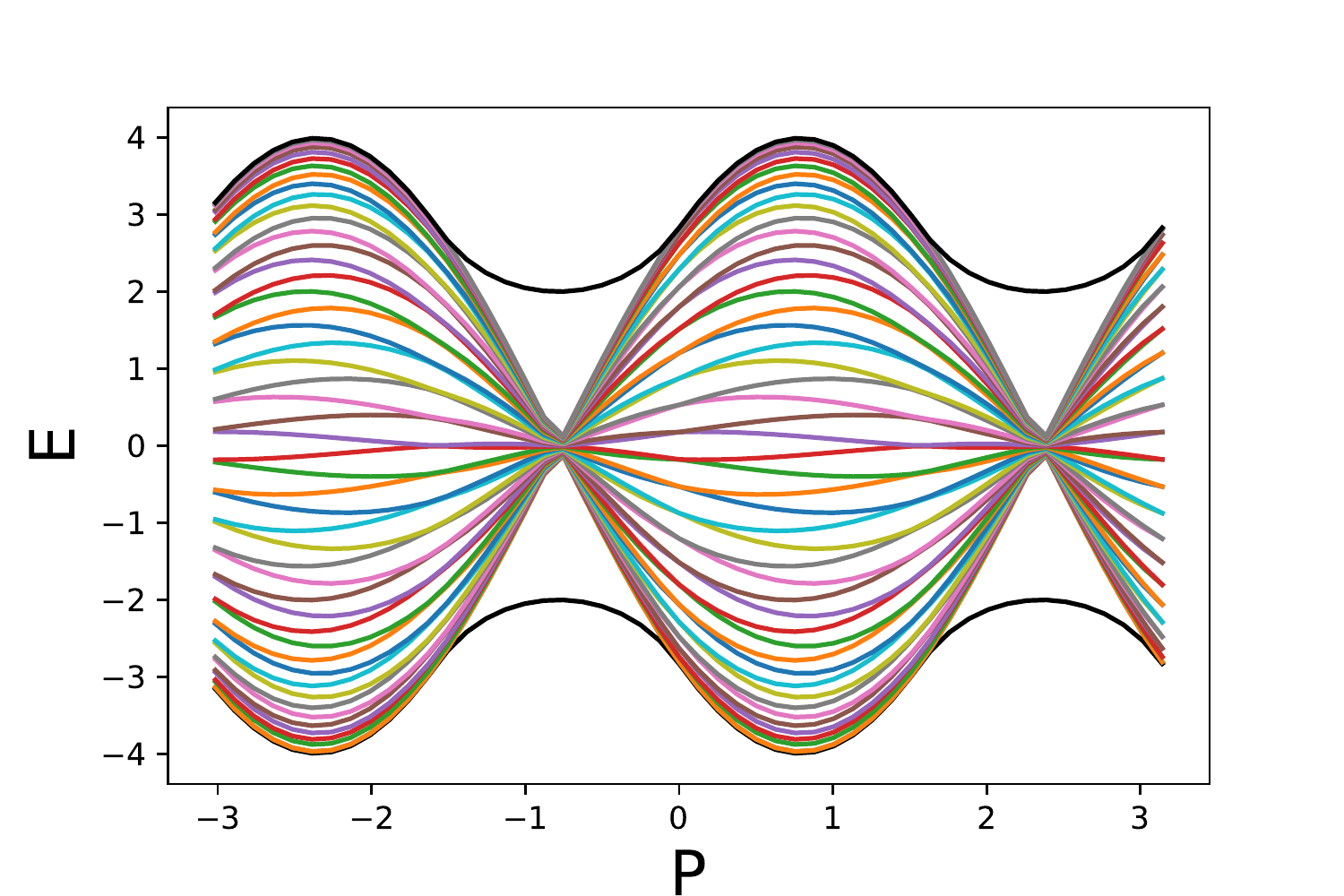}}
\\
\subfigure[]{\includegraphics[width=5.9cm]{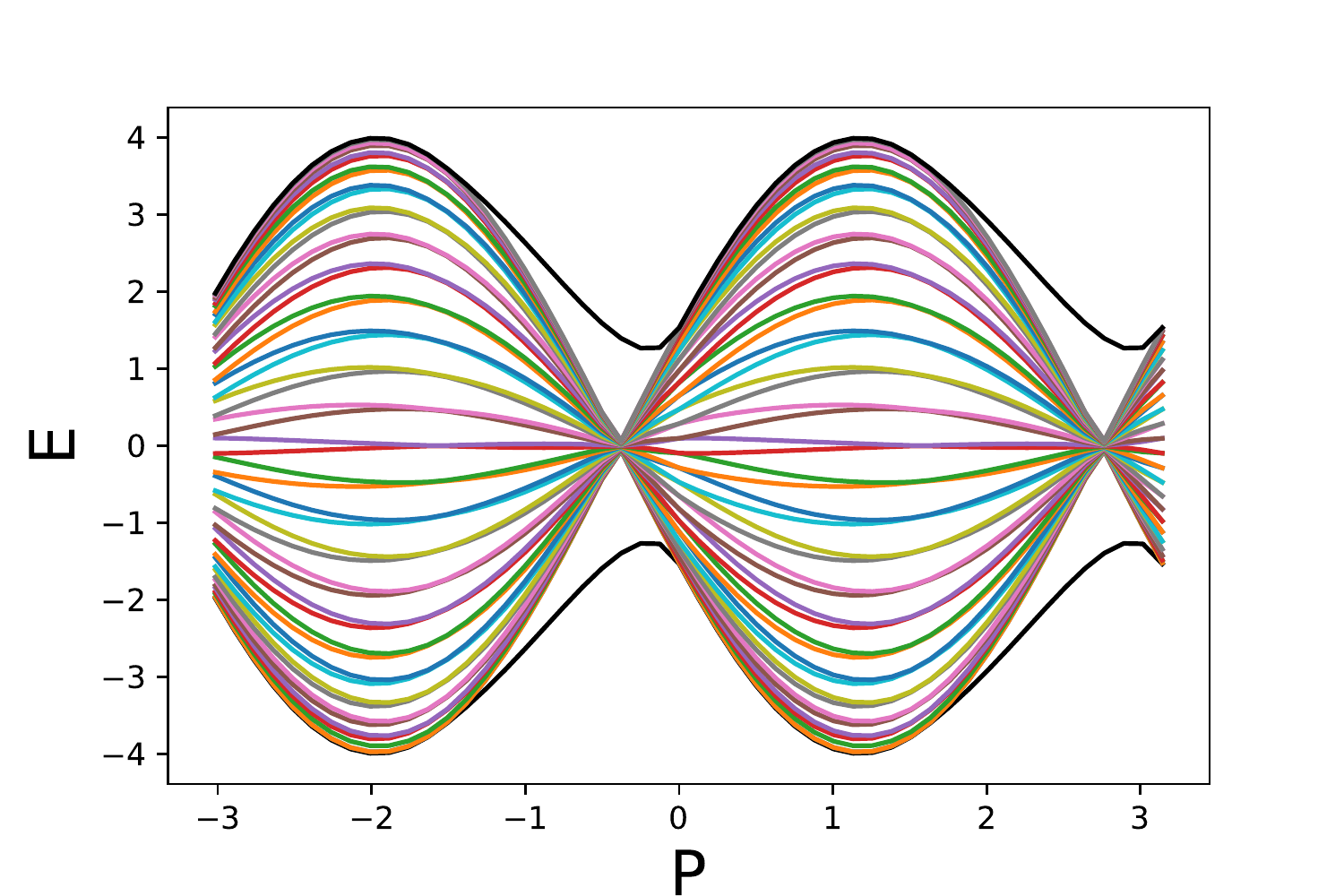}}
\subfigure[]{\includegraphics[width=5.9cm]{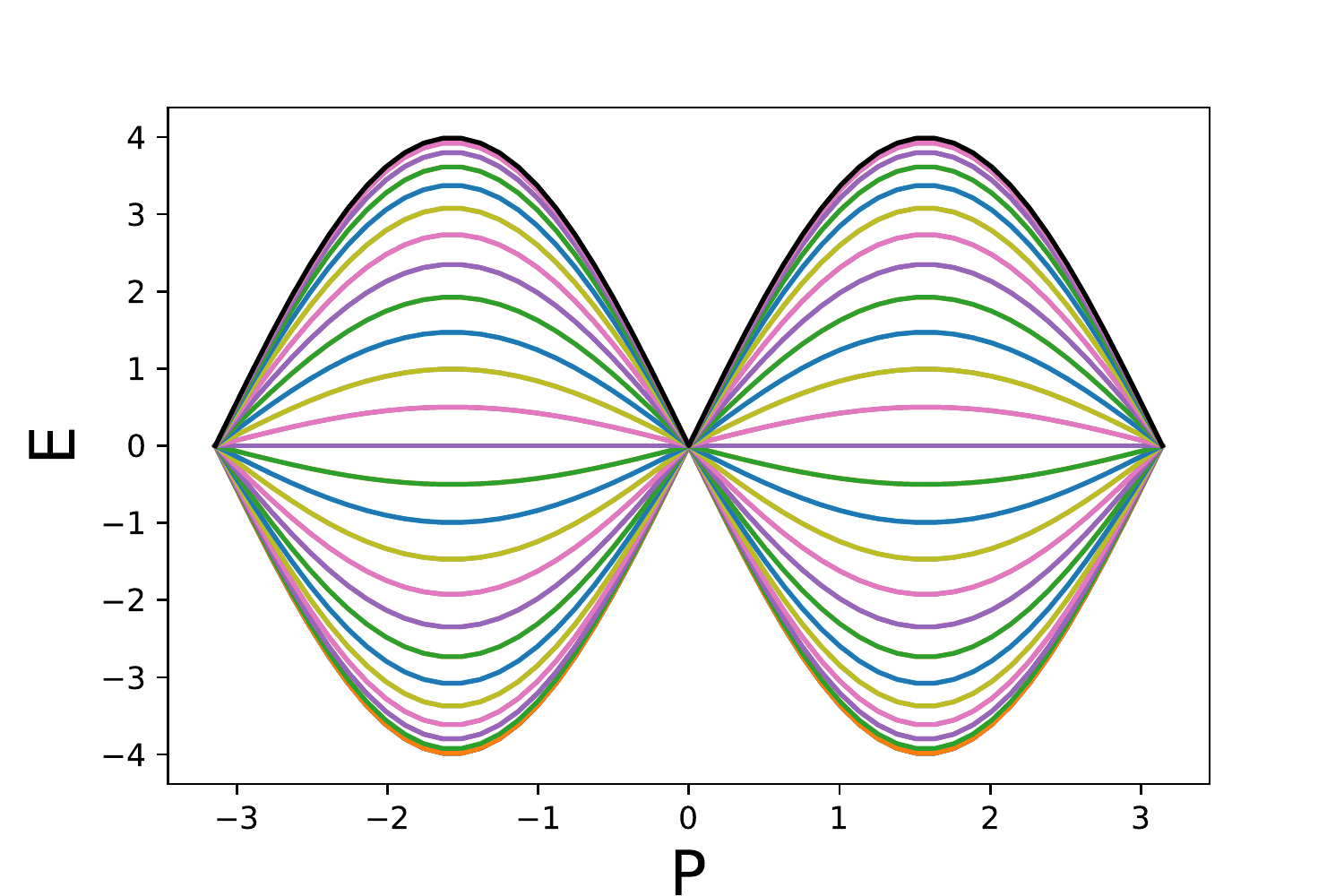}}
\subfigure[]{\includegraphics[width=5.9cm]{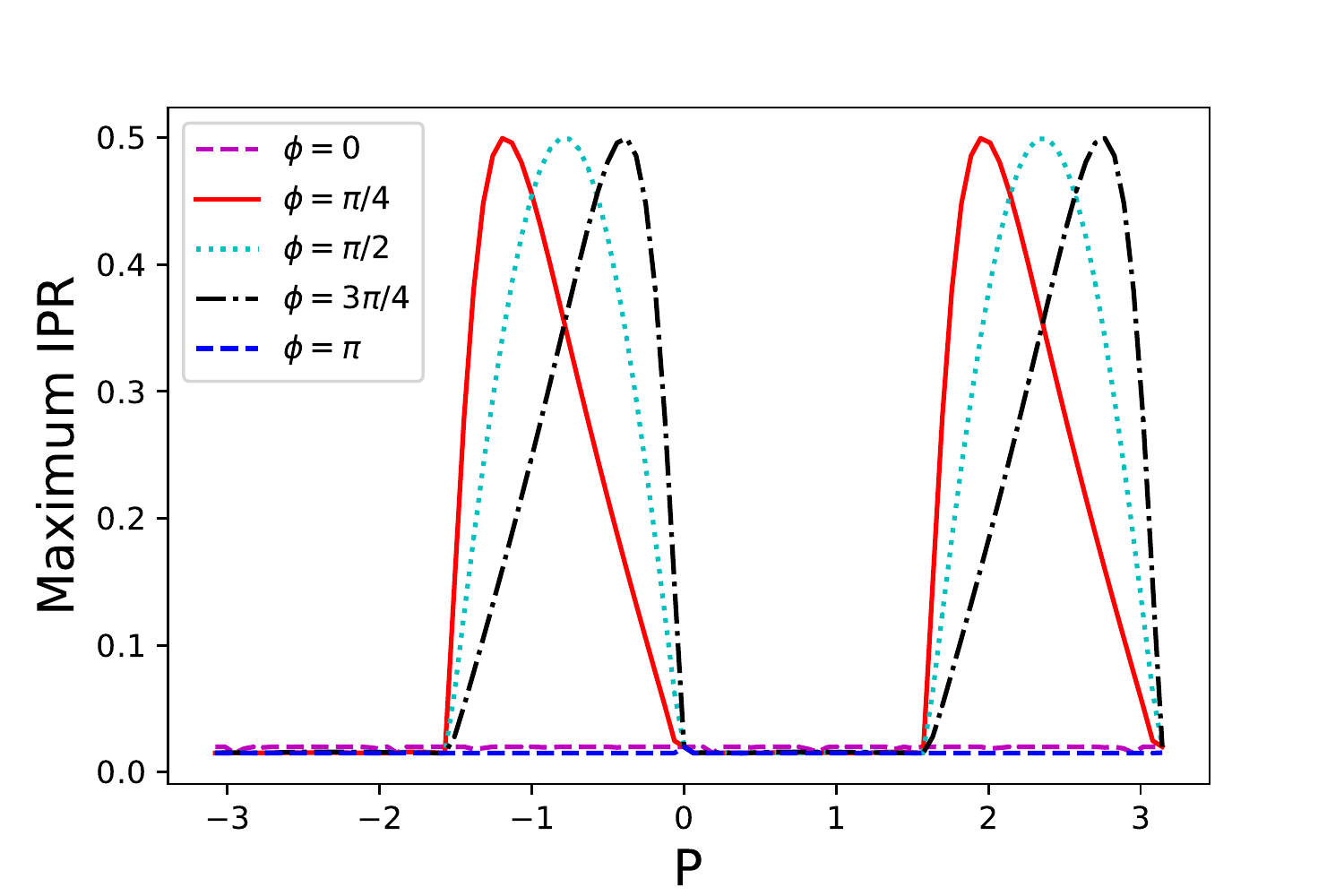}}
\caption{(a-e) Energy levels of two-particle states vs the center-of-mass 
momentum $P$ for $\phi = 0, ~\pi/4, ~\pi/2, ~3\pi/4$ and $\pi$, found 
numerically for a relative coordinate system with 50 sites, $t_2 = -1$ and 
$\mu = 0$. The isolated lines (black solid) correspond to two-particle bound 
states, while all the other lines correspond to the two-particle 
continuum of states. 
There are no bound states for $\phi = 0$ and $\pi$. 
(f) Maximum IPR vs $P$ for the same values of $\phi$ for a relative 
coordinate system with 100 sites. Whenever the maximum IPR is nonzero (for 
an infinitely large system), it corresponds to a two-particle bound state.}
\label{sa1fig05} \end{figure} \end{center}
\end{widetext}

Figures~\ref{sa1fig05} (a-e) shows the energy levels of a system with 50 sites 
(for the relative coordinate problem) as a function of $P$ for various values
of $\phi$, with $t_2 = -1$. In each of the figures, we see a continuum 
of states lying in the range $[-|4 \cos (P - \phi/2)|, |4 \cos (P - \phi/2)|]$,
in agreement with the discussion above. We also see pairs of bound 
states with opposite energies given by the isolated black solid lines; these
appear in the regions given by Eq.~\eqref{boundcond}. Figure~\ref{sa1fig05} 
(f) shows the maximum value of the inverse participation ratio (IPR)
as a function of $P$ for different values of $\phi$. Given a normalized
eigenstate $\psi_a (n_2 - n_1)$ of the Hamiltonian, the IPR is defined
as $\sum_{n_2 - n_1} | \psi_a (n_2 - n_1)|^4$. It is known that this is
a useful diagnostic for the presence of bound states. As the system
size is taken to infinity, the IPRs of extended states (whose energies
form a continuum) go to zero, while the IPRs of localized states 
(i.e., bound states) remain finite. We indeed see that the ranges of $P$
in Fig.~\ref{sa1fig05} (f) where the maximum value of the IPR is large 
coincides precisely with the ranges in Figs.~\ref{sa1fig05} (a-e) where
there are bound states.

It is interesting to consider what happens if we shift 
$P \to P + \pi$; this changes the center-of-mass wave function by a factor of 
$e^{i \pi (n_1 + n_2)/2}$. We then see from the discussion following
Eq.~\eqref{fn} that all the hoppings of the relative coordinate problem 
flip sign. We then find that the energies of both continuum and bound states
remain the same but the relative coordinate wave functions change
by a factor of $e^{i \pi (n_2 - n_1)/2}$ (we recall that $n_2 - n_1$ can
only change by multiples of 2 when the particle hop). Combining the changes 
in the center-of-mass and relative coordinate wave functions, we see that the
total wave function changes by $e^{i \pi n_2}$ which is equal to $+1$ 
since $n_2$ is even. We therefore see 
that the wave functions remain the same for all values of $n_1$ and $n_2$.
Thus the energy spectrum and eigenstates do not change if $P$ is shifted by 
$\pi$. Both the energy levels and IPR values shown in Figure~\ref{sa1fig05} 
are consistent with this observation.

\subsection{Implications for low-density limit}
\label{sec5a}

The existence of two-particle bound states may have significant implications
for the nature of the ground state in the thermodynamic limit, i.e., the
limit in which the number of particles $N$ and the number of sites $L$ are 
both taken to infinity, keeping the particle density 
$\rho = N/L = 1/2 + \eta$ fixed. 
For simplicity, we consider the low-density limit where $\rho \ll 1$, so that 
interactions between more than two particles can be ignored (hence we are 
ignoring the possibility of bound states of three or more particles). In this 
limit, it may be preferable for pairs of particles to 
occupy the two-particle bound states with negative energy (which lie below
the two-particle continuum as shown in Fig.~\ref{sa1fig05}) rather than for 
the particles to occupy single-particle states independently of each other. 
We will now briefly examine the values of density and $\phi$ where this 
is likely to happen.

In the low-density limit, pairs of particles will only occupy states near the
minima of the energy levels shown in Eq.~\eqref{epk}, namely, near $P - \phi/2 
= 0$ and $k =0$ or near $P - \phi/2 = \pi$ and $k = \pi$, i.e., near
$P = \phi/2$ or $\pi + \phi/2$. On the other hand, 
Eq.~\eqref{boundcond} shows that bound states can appear only if $-\pi/2 < 
P < 0$ or $\pi/2 < P < \pi$. Thus $P$ must deviate from the minima at
$\phi/2$ or $\pi + \phi/2$ by at least $-\phi/2$ for bound states to start
appearing. As a result, the single-particle momenta $k_1$ and $k_2$ must
deviate from their minimum possible value by $-\phi/4$, following 
Eq.~\eqref{k1k2}. Now, since particles
on a particular sublattice can only move in multiple of two sites, the ranges 
of $k_1$ and $k_2$ are equal to $\pi$ and they are quantized in units of 
$2 \pi/L$ (hence each of them can take $L/2$ values). Thus a deviation of
$\phi/4$ from the minimum of the energy 
means that the system must have at least $(\phi/4)/(2\pi/L) = \phi L
/(8 \pi)$ particles on each sublattice occupying the range of $\phi/4$
near $P=0$ and an equal number of particles occupying the same range near
$P=\pi$. Hence the total number of particles must be equal to at least
$\phi L/(2 \pi)$, implying that the particle density must be at least $\phi /
(2 \pi)$ before bound states start appearing in the ground state. To be 
consistent with the low-density limit, we see that $\phi$ should be much 
smaller than $\pi$. We therefore see that if $\phi$ is small, we require
the density to be of the order of $\phi/(2 \pi)$ before bound states can start
playing a role in the ground states of the system. When the density is
larger than this amount, we may have to re-analyze the mean field
theory done in Sec.~\ref{sec3} to take the bound states into account.

In conclusion, the possible effects of two-particle bound states on the 
ground state may be an interesting problem for detailed studies in the future.

\section{Discussion}
\label{sec6}

We first summarize our results. We have studied a one-dimensional
model of spinless fermions in which particles have only next-nearest-neighbor 
hoppings, where the phase of the hopping depends on a statistical phase
$\phi$ and the number of fermions (0 or 1) on the intermediate site. (This 
model is related, by a unitary transformation, to a model of particles which 
satisfy a generalized statistics which is governed by the parameter $\phi$).
This kind of hopping leads to four-fermion interactions between particles
living on the even and odd sublattices. We looked at the properties of the 
model under particle-hole, parity and time-reversal
transformations. We find that the model is not invariant under $P$ and 
$T$ separately but is invariant under the product $PT$.
We then studied a mean field theory of the model and found that, for a range
of values of the chemical potential, there are four Fermi points; the
locations of these points depend on $\phi$ and the filling which is described 
by a parameter $\eta$. The Fermi points correspond to two right-moving and 
two left-moving points.

We then developed a bosonized theory of the excitations involving modes near 
the Fermi points; this theory involves four bosonic fields. We find that the 
theory has nontrivial interactions only if $\phi \ne 0$ and, more 
remarkably, only if we are away from half-filling (i.e., if $\eta \ne 0$).
The original fermionic theory turns out to lead to 
a variety of terms in the bosonic language. Some of the terms are quadratic
in the bosonic fields while others involve cosines of linear
combinations of the bosonic fields. We diagonalized the quadratic terms using 
Bogoliubov transformations. It turns out that the four bosonic fields
decouple into two sets of pairs of bosonic fields, thus giving rise to two 
separate Tomonaga-Luttinger liquids with different Luttinger parameters 
$K_1$ and $K_2$
and different velocities. (The right- and left-moving bosonic fields turn 
out to have different velocities because of the lack of parity symmetry). In 
terms of these parameters we found the scaling dimensions of the cosine terms 
mentioned above and the regimes of parameters $\phi$ and $\eta$ where they
are relevant or irrelevant. Based on these scaling dimensions and RG flow 
arguments to first order in the couplings, we found that in one of the 
Tomonaga-Luttinger liquids, one of the couplings may grow at long distances
and may thereby produce a gap. However, we need to examine
the effects of higher order terms in the RG equations to understand if this
really occurs. The other Tomonaga-Luttinger liquid always remains gapless.
Next, we calculated the correlation functions of
the twelve different charge density and superconducting order parameters 
that exist in this model, and found that they all decay as power laws. As a 
function of $\phi$ and $\eta$, we found which of these order parameters is 
the most relevant (i.e., has the smallest scaling dimension) and therefore 
will dominate the correlations in the long-distance limit. We emphasize that 
exactly at half-filling ($\eta = 0$), the system is noninteracting for all 
values of $\phi$ and is described by two Tomonaga-Luttinger liquids which 
form a conformal field theory with $c=2$.

Finally, we studied the energy spectrum 
of two particles, one living on each sublattice. We found that there can
be a bound state of the two particles depending on the value of their
center-of-mass momentum $P$ and $\phi$. Interestingly, the energies of some 
of the bound states, when they exist, lie below the two-particle continuum.
This implies that these bound states can play a role in the form of the
ground state in the thermodynamic limit, and we made an estimate of the
minimum particle density when this might occur.

We now list some problems which may be useful to study in the future.

\noi (i) In the second block of the two Tomonaga-Luttinger liquids, the RG 
equations for $\lm_1, ~\lm_2, ~\lm_3$ and $K_2$ need to be found up to 
second order to obtain a better understanding of the fixed point 
that the system may reach at long distances~\cite{cardy,affleck,dutta,giam}.
In particular, we would like to know if the fixed point is gapless or gapped.

\noi (ii) It may be interesting to examine what happens if the fillings in 
the even and odd sublattice are not the same. This would require us to 
take the chemical potentials to be different on the two sublattices in order
to develop a mean field theory followed by bosonization.

\noi (iii) The effect of the two-particle bound states on the ground state 
of the system in the thermodynamic limit needs to be 
understood~\cite{vidal,eckholt1}. For example, we can investigate if the 
ground state is a condensate of pairs of particles.

\noi (iv) We can study if there are bound states of three or more particles 
for some values of the center-of-mass momentum and $\phi$.

\noi (v) We may ask what happens when a nearest-neighbor hopping 
$t_1$ is present in addition to the next-nearest-neighbor hopping $t_2$ 
(see Eq.~\eqref{ham2}). Such a model is considerably more complicated 
to analyze since it does not conserve the number of particles on the two 
sublattices separately, and the models with filling fractions $1/2 + \eta$
and $1/2 - \eta$ are no longer related to each other by a particle-hole
transformation. We find that for small values of $t_1 /t_2$, 
there is no significant change in the results, either for the bosonization
analysis or the two-particle bound states, compared to the results for
$t_1 = 0$ that we have presented in this paper. For $|t_1 / t_2| > 2$,
however, a different phase appears in which there are only two Fermi points.
This phase has been studied in detail at half-filling in 
Ref.~\onlinecite{agarwala1}.

We conclude by discussing possible realizations of the model considered in 
this paper. Apart from theoretical ideas for realizing generalized statistics 
in one dimension~\cite{keil,strater,gres,card,gres2}, systems of fermionic or 
bosonic atoms with density-dependent hoppings have been proposed 
theoretically~\cite{eckholt2,itin,chhaj,ghosh,liberto,agarwala2,hudomal,stas,
gotta} and realized experimentally~\cite{meinert,gorg}. The system studied in
Ref.~\onlinecite{gorg} is particularly promising since the phase of the
hopping of spin-1/2 fermions in one spin state is dependent on the density of
fermions in the opposite spin state, analogous to our model where sublattice
plays the role of spin.

\vspace{.8cm}
\centerline{\bf Acknowledgments}
\vspace{.5cm}

The authors thank Adhip Agarwala and Subhro Bhattacharjee for many useful
discussions. S.A. thanks MHRD, India for financial support through a PMRF.
D.S. thanks DST, India for Project No. SR/S2/JCB-44/2010 for financial support.

\vspace{.5cm}
\appendix

\section{Bogoliubov transformation of bosons with opposite chiralities and 
unequal velocities}

In this Appendix, we will discuss the Bogoliubov transformation which was 
used to diagonalize the Hamiltonians in Eqs.~\eqref{ham14} and \eqref{hamx}.
We consider a model with two bosonic fields with opposite chiralities 
and unequal velocities $v_{1}$ and $v_{2}$ and a coupling $\lambda$ between 
them. The Hamiltonian of this system is given by
\begin{eqnarray}
H &=& \sum_{q>0} ~q ~[v_{1}b_{q,R}^{\da}b_{q,R}+v_{2}b_{q,L}^{\da}b_{q,L} 
\non \\
&& ~~~~~~~~~~+\lambda (b_{q,R}^{\da}b_{q,L}^{\da}+b_{q,L}b_{q,R})], 
\label{hamapp1} \end{eqnarray}
where $b_{q}$ and $b_{q}^{\da}$ are bosonic annihilation and creation 
operators which satisfy 
\begin{eqnarray} \left[b_{q,\nu},b_{q',\nu'}^{\da}\right]&=&
\delta_{q,q'}\delta_{\nu,\nu'}, \non \\
\left[b_{q,\nu},b_{q',\nu'}\right]&=& 0, \non \\
\left[b_{q,\nu}^{\da},b_{q',\nu'}^{\da}\right] &=& 0, \label{comm} 
\end{eqnarray}
where $\nu, ~\nu' = R,L$.
We will discuss the diagonalization of the Hamiltonian in Eq.~\eqref{hamapp1}
by a Bogoliubov transformation for a particular value of $q$. The 
Bogoliubov transformation is given by
\begin{eqnarray}
b_{q,R}&=&\alpha ~\Tilde{b}_{q,R} ~+~ \beta ~\Tilde{b}_{q,L}^{\da},\non\\
b_{q,L}&=&\alpha ~\Tilde{b}_{q,L} ~+~ \beta ~\Tilde{b}_{q,R}^{\da},\non\\
\alpha &=&\cosh \theta, ~~~~~ \beta ~=~ \sinh \theta. \end{eqnarray}
We have chosen $\alpha$ and $\beta$ to have these forms to satisfy the 
commutation relations given in Eq~\eqref{comm} for the $\tilde b$ 
operators also. The Hamiltonian for a particular $q$ is then given by
\begin{widetext}
\begin{eqnarray}
H&=&q\left(\frac{v_{1}+v_{2}}{2}\left(\alpha^2+\beta^2\right)+ 2\alpha\beta
\lambda\right)\left(\Tilde{b}_{q,R}^{\da}b_{q,R}+ \Tilde{b}_{q,L}^{\da}
\Tilde{b}_{q,L}\right) ~+~ q ~\frac{v_{1}-v_{2}}{2} \left(\Tilde{b}_{q,R}^{\da}
b_{q,R}-\Tilde{b}_{q,L}^{\da}\Tilde{b}_{q,L} \right)\non\\
&& + ~q\left[(v_{1}+v_{2})\alpha\beta+\lambda\left(\alpha^2+\beta^2\right)
\right]\left(\Tilde{b}_{q,R}^{\da}\Tilde{b}_{q,L}^{\da}+\Tilde{b}_{q,L}b_{q,R}
\right). \label{hamapp2} \end{eqnarray}
\end{widetext}
To have a diagonal Hamiltonian, $\alpha$ and $\beta$ must satisfy
$(v_{1}+v_{2})\alpha\beta+\lambda (\alpha^2+\beta^2) = 0$ which implies
\begin{equation} \tanh(2\theta)=-\frac{2\lambda}{v_{1}+v_{2}}. \end{equation}
Using this in Eq.~\eqref{hamapp2}, we obtain 
\begin{eqnarray} H&=&q\left(v ~+~ \frac{v_{1}-v_{2}}{2}\right)
\Tilde{b}_{q,R}^{\da}b_{q,R} \non\\
&&+ ~q\left(v-\frac{v_{1}-v_{2}}{2}\right)\Tilde{b}_{q,L}^{\da}\Tilde{b}_{q,L},
\non\\
v&=&\frac{v_{1}+v_{2}}{2}\sqrt{1-\frac{4\lambda^2}{(v_{1}+v_{2})^2 }}.
\label{hamapp3} \end{eqnarray}
From Eq.~\eqref{hamapp3}, we see that the system has a well-defined ground 
state if $v$ is real and larger than $|v_1 - v_2|/2$. We find that these 
conditions hold if
\begin{equation} v_{1}v_{2} ~>~ \lambda^2. \end{equation}
The new bosonic fields have the forms
\begin{eqnarray} \Tilde{b}_{q,R}&=&\frac{b_{q,R}+\gamma\: 
b_{q,L}^{\da}}{\sqrt{1-\gamma^2}}, \non\\
\Tilde{b}_{q,L}&=&\frac{b_{q,L}+\gamma\: b_{q,R}^{\da}}{\sqrt{1-\gamma^2}},
\non\\
\gamma&=&\frac{1-K}{1+K}, \end{eqnarray}
where
\begin{equation} K=\sqrt{\frac{v_{1}+v_{2}-2\lambda}{v_{1}+v_{2}+2\lambda}}.
\label{k} \end{equation}
We note that the parameters $v$ and $K$ in Eqs.~\eqref{hamapp3} and \eqref{k}
do not depend on the value of $q$.

\section{Mapping between $\phi$ and $\pi + \phi$}

In this Appendix, we will show that the systems defined by Eq.~\eqref{ham4}
for $\phi$ and $\pi + \phi$ can be mapped to each other by transforming the
fermionic operators in a particular way.
We will first consider an infinite system since the transformation is
easier to discuss in that case. We consider the Hamiltonian $H$ given in 
Eq.~\eqref{ham10} which we rewrite as
\bea H &=& \sum_{j} ~[\cos(\phi/2) ~(c_{j}^\da c_{j+2}+c_{j+2}^\da c_{j}) 
\non \\
&& ~~~~~~~+i \sin(\phi/2) ~(2 n_{j+1} -1) ~(c_{j}^\da c_{j+2}-c_{j+2}^\da 
c_{j}) \non \\
&& ~~~~~~~-\mu ~c_{j}^\da c_{j}]. \label{ham19} \eea
We now use the fact that $2 n_j - 1$ is a Hermitian operator with eigenvalues
equal to $\pm 1$; further, it anticommutes with $c_j$ and $c_j^\da$ but 
commutes with $c_k$ and $c_k^\da$ for all $k \ne j$. We define new fermionic 
operators
\bea {\tilde c}_j &=& c_j ~\prod_{l=0}^\infty ~(2 n_{j+1+2l} ~-~ 1) 
~~~~{\rm if}~~~ j ~~~{\rm is~ even}, \non \\
{\tilde c}_j &=& c_j ~\prod_{l=0}^\infty ~(2 n_{j-1-2l} ~-~ 1) 
~~~~{\rm if}~~~ j ~~~{\rm is~ odd}. \label{ctilde} \eea
In words, ${\tilde c}_j$ is equal to $c_j$ multiplied by a string of 
$2 n_k - 1$ on its right on all the sites of the odd sublattice if $j$ lies
on the even sublattice and by a string of $2 n_k - 1$ on its left on all the 
sites of the even sublattice if $j$ lies on the odd sublattice.
The crucial point to note is that the transformations in Eq.~\eqref{ctilde}
maintain the anticommutation relations $\{ {\tilde c}_j, {\tilde c}_k \} = 0$ 
and $\{ {\tilde c}_j, {\tilde c}_k^\da \} = \de_{jk}$ for all values of 
$j, ~k$, and ${\tilde c}_{j}^\da {\tilde c}_{j} = c_j^\da c_j = n_j$. In terms 
of the new operators, the Hamiltonian in Eq.~\eqref{ham19} takes the form
\bea H &=& \sum_{j} ~[\cos(\phi/2) ~(2 n_{j+1} -1) ~({\tilde c}_{j}^\da 
{\tilde c}_{j+2}+ {\tilde c}_{j+2}^\da {\tilde c}_{j}) \non \\
&& ~~~~~~~+i \sin(\phi/2) ~({\tilde c}_{j}^\da {\tilde c}_{j+2}-
{\tilde c}_{j+2}^\da {\tilde c}_{j}) \non \\
&& ~~~~~~~-\mu ~{\tilde c}_{j}^\da {\tilde c}_{j}]. \label{ham20} \eea

Next, we do another transformation
\beq {\tilde c}_j ~\to~ e^{ij \pi/4} ~{\tilde c}_j ~~~{\rm and}~~~
{\tilde c}_j^\da ~\to~ e^{-ij \pi/4} ~{\tilde c}_j^\da. \label{ijpi} \eeq 
Then Eq.~\eqref{ham20} turns into
\bea H &=& \sum_{j} ~[- \sin(\phi/2) ~({\tilde c}_{j}^\da {\tilde c}_{j+2}
+{\tilde c}_{j+2}^\da {\tilde c}_{j}) \non \\
&& ~~~~~~~+i \cos(\phi/2) ~(2 n_{j+1} -1) ~({\tilde c}_{j}^\da {\tilde c}_{j+2}
-{\tilde c}_{j+2}^\da {\tilde c}_{j}) \non \\
&& ~~~~~~~-\mu ~{\tilde c}_{j}^\da {\tilde c}_{j}]. \label{ham21} \eea
Comparing Eqs.~\eqref{ham19} and \eqref{ham21} we see that $\phi$ has 
effectively changed to $\pi + \phi$ so that $\cos (\phi/2) \to - 
\sin (\phi/2)$ and $\sin (\phi/2) \to \cos (\phi/2)$.

We now discuss how the above transformations work for a finite-sized system
with periodic boundary conditions.
We assume that the total number of sites $N$ is even so that each sublattice 
has $N/2$ sites and the site indices in Eq.~\eqref{ham19} can only go from 
1 to $N$. Then the string of $2n_k -1$ in the first line of 
Eq.~\eqref{ctilde} ends on the right at $k=N-1$ and the string in the second
line ends on the left at $k=2$. Then if we look at the hopping between sites 
1 and $N-1$ or between 2 and $N$, we find that they will satisfy periodic
boundary conditions only if 
\beq 
\prod_{l=1}^{N/2} ~(2 n_{2l-1} ~-~ 1) ~=~ 1 ~~~{\rm and}~~~ 
\prod_{l=1}^{N/2} ~(2 n_{2l} ~-~ 1) ~=~ 1. \eeq
These conditions imply that the number of unoccupied sites (which have
$n_k = 0$) must be even on both even and odd sublattices, namely, $N/2$ 
minus the number of particles must be even on both sublattices.
Next, we see that the transformation in Eq.~\eqref{ijpi} will satisfy
periodic boundary conditions if $e^{iN\pi/4} = 1$, i.e., if $N$ is a 
multiple of 8. Hence $N/2$ is an even number and therefore
the previous condition implies that the number of particles on each sublattice
should be even so that the mapping from $\phi$ to $\pi + \phi$ can
work with periodic boundary conditions.

We note that the transformations given in Eq.~\eqref{ctilde} between the old 
and new fermionic operators are highly nonlocal. Perhaps for this reason, 
the symmetry between $\phi$ and $\pi + \phi$ is not evident in the results
obtained by bosonization. Namely, the expressions for various
quantities in Sec.~\ref{sec4}, such as $\theta$, $v$, $K_1$ and $K_2$ in 
Eqs.~\eqref{thetamu}, \eqref{vel}, \eqref{vK1} and \eqref{vK2}, are 
not invariant under $\phi \to \pi + \phi$.

\end{document}